\documentclass[manuscript]{aastex}
\usepackage{emulateapj5}

\shorttitle{DA\,495 -- an aging PWN}
\shortauthors{Kothes et al.}

\begin{document}

\title{DA\,495 -- an aging pulsar wind nebula}

\author{R. Kothes\altaffilmark{1,2}, T.L. Landecker\altaffilmark{1},
W. Reich\altaffilmark{3}, S. Safi-Harb\altaffilmark{4}, 
Z. Arzoumanian\altaffilmark{5}}
\altaffiltext{1}{National Research Council of Canada,
              Herzberg Institute of Astrophysics,
              Dominion Radio Astrophysical Observatory,
              P.O. Box 248, Penticton, British Columbia, V2A 6J9, Canada}

\altaffiltext{2}{Department of Physics and Astronomy, University of Calgary,
             2500 University Drive N.W., Calgary, AB, Canada}

\altaffiltext{3}{Max-Planck-Institut f\"ur Radioastronomie, Auf dem H\"ugel
             69, 53121 Bonn, Germany}

\altaffiltext{4}{Canada Research Chair, Department of Physics and Astronomy, 
University of Manitoba, Winnipeg, MB, R3T 2N2, Canada}

\altaffiltext{5}{CRESST and X-ray Astrophysics Laboratory, NASA-GSFC, 
Greenbelt MD 20771, U.S.A.}

\email{roland.kothes@nrc-cnrc.gc.ca, tom.landecker@nrc-cnrc.gc.ca,
wreich@mpifr-bonn.mpg.de, safiharb@cc.umanitoba.ca, 
zaven@milkyway.gsfc.nasa.gov}

\begin{abstract}
We present a radio continuum study of the 
pulsar wind nebula (PWN) DA\,495 (G65.7+1.2), including images of total 
intensity and linear
polarization from 408 to 10550\,MHz based on the Canadian Galactic
Plane Survey and observations with the Effelsberg 100-m Radio
Telescope. Removal of flux density contributions from a superimposed
\ion{H}{2} region and from compact extragalactic sources reveals a
break in the spectrum of DA\,495 at 1.3\,GHz, with a spectral index
${\alpha}={-0.45~\pm~0.20}$ below the break and
${\alpha}={-0.87~\pm~0.10}$ above it (${S}_\nu \propto{\nu^{\alpha}}$).  The
spectral break is more than three times lower in frequency than the
lowest break detected in any other PWN. The break in the
spectrum is likely the result of synchrotron cooling, and DA\,495, at
an age of $\sim$20,000 yr, may have evolved from an object similar to
the Vela~X nebula, with a similarly energetic pulsar. We find a 
magnetic field of $\sim$1.3~mG inside the nebula. After correcting for the 
resulting high internal rotation measure, the magnetic field structure is 
quite
simple, resembling the inner part of a dipole field projected onto the plane 
of the sky,
although a toroidal component is likely also present. The dipole field
axis, which should be parallel to the spin axis of the putative pulsar,
lies at an angle of ${\sim}50\degr$ east of the North Celestial Pole
and is pointing away from us towards the south-west. The upper limit 
for the radio surface brightness of any shell-type supernova remnant
emission around DA\,495 is 
$\Sigma_{1\,GHz} \sim 5.4 \times 10^{-23}$~Watt~m$^{-2}$~Hz$^{-1}$~sr$^{-1}$
(assuming a radio spectral index of $\alpha = -0.5$),
lower than the faintest shell-type remnant known to date.
\end{abstract}


\section{Introduction}

The Crab Nebula, the remnant of the supernova explosion of 1054, is an
astrophysical laboratory of immense value where we can study the
development of a young pulsar and the synchrotron nebula generated by
the particles that it injects. The Crab Nebula has become the
archetype of a class of objects earlier known as filled-center,
plerionic, or Crab-like supernova remnants (SNRs), but now called
pulsar wind nebulae (PWNe) in recognition of their energy source. PWNe
are a minority in SNR catalogs, less than 10\% in number, but if we
include the composite SNRs, synchrotron nebulae centrally placed
within the more usual SNR shell, their ranks swell to 15\% of the
Galactic SNR catalog (Green 2004).

PWNe are characterized by a flat radio spectral index, $\alpha$, in
the range ${0} > {\alpha} > {-0.3}$ (where the flux density $S$ at
frequency $\nu$ varies as ${S} \propto {{\nu}^{\alpha}}$) and a
central concentration of emission gradually declining to the outer
edge. Their appearance reflects the central energy source and their
radio spectrum reflects the energy spectrum of the injected
particles. The more common shell SNRs exhibit steep outer edges and
steep radio spectra (generally ${\alpha} < {-0.3}$).

We have historical records of fewer than ten supernova events
(Stephenson \& Green 2002).  By chance two of those have left
PWNe, the Crab Nebula and 3C 58, with no detectable trace of
accompanying shells, and one a composite remnant, G11.2$-$0.3. Thus we
have three examples in which we can closely follow the evolution of
young PWNe, but the relatively small number of known
PWNe has provided little observational basis for our
understanding of the evolution of these objects through later
stages. This paper presents and analyzes new observations of DA\,495
(G65.7+1.2), which seems likely to be a PWN of advanced age.

DA\,495, while an outlier among PWNe, is without doubt a member
of that class. Its radio appearance, first seen in the high-resolution
images of \citet{land83}, has the signature of a PWN, a central
concentration of emission smoothly declining with increasing radius,
with no evidence of a surrounding shell (although there is a
depression in intensity near its center). The only indication that it might
not be a PWN is its measured radio spectral index, ${\alpha}
\approx {-0.5}$ \citep{koth06b}, much steeper than is usual for this
class of objects, and more closely resembling that of a shell-type
SNR. Nevertheless, shell remnants have steep outer edges where kinetic
energy is being converted to radio emission near the shock front,
while in DA\,495 the emissivity distribution is centrally concentrated
\citep[Fig.~3]{land83} indicating that the energy source is
central. Nevertheless, on the basis of the central depression in
intensity, \citet{velu89} argue that DA\,495 might be a shell remnant
with an exceptionally thick shell formed by a reverse shock
interacting with an unusual amount of slowly moving ejecta.

Recent analysis of X-ray data for DA\,495, obtained from the ROSAT and
ASCA archives, dramatically confirms the PWN interpretation of
this object by revealing a compact central object surrounded by an
extended non-thermal X-ray region \citep{arzo04}. Analysis of Chandra
data \citep{arzo08} shows a non-thermal X-ray nebula of extent
$\sim$30\arcsec\ surrounding the central compact object. The compact
source is presumably the pulsar (required to power the
synchrotron nebula), and we shall refer to it as such (even though
pulsed emission has not been detected). In this paper we present new
radio images of DA\,495 which include linear polarization
information. Analysis of high-resolution radio images allows us to
remove superimposed sources. After this the integrated spectrum of the
source exhibits a spectral break characteristic of a synchrotron
nebula, and the observed parameters allow an estimate of the strength
of the magnetic field within the nebula. Polarization images reveal
the structure of the field within this PWN.

\section{Observations and Data Analysis}

The DRAO observations at 408 and 1420 MHz are part of the Canadian
Galactic Plane Survey \citep[CGPS,][]{tayl03}. Data from single-antenna
telescopes were incorporated with the Synthesis
Telescope data to give complete coverage of all structures down to the
resolution limit \citep[details of the process can be found in][]{tayl03}. 
The one exception is the polarization data at 1420~MHz,
for which no single-antenna data were available. However, the
telescope is sensitive to all structures from the resolution limit
(${\sim}1'$) up to $45'$. The absence of single-antenna data is not a
concern for observations of DA\,495, whose full extent is $\le 30'$.
The CGPS polarization data were obtained in four bands (1407.0, 1413.0, 1427.6,
and 1434.5~MHz) each of width 7.5~MHz, allowing derivation of 
rotation measures (RMs) between those frequencies.

The Effelsberg observations were carried out well before this
investigation began, and were retrieved from archived
data. Observations at 4850 and 10550 MHz were made at different times,
but in each case in excellent weather.  The receiver systems,
installed at the secondary focus of the telescope, received both hands
of circular polarization, with two feeds at 4850 MHz and four feeds at
10550 MHz. Standard data-reduction software based on the NOD2 package
\citep{hasl82} was employed throughout.  Data reduction relied on
``software beam switching'' to reject atmospheric emission above the
telescope \citep{mors86}. Here every feed is equipped with its
own amplifier to allow parallel recording of data. For each pair of
feeds a difference map is calculated in which disturbing atmospheric
effects are largely subtracted.  These dual-beam maps are restored to
the equivalent single-beam map with the algorithm described by
\citet{emer79}. To reduce scanning effects, multiple maps -- coverages
-- of the field were observed at different parallactic angles. These
scanning effects arise from instrumental offsets and drifts that
produce different baselevels for each individual scan, a typical
result from raster scanning. Individual coverages were combined using
the ``plait'' algorithm \citep{emer88}, by destriping the maps in the
Fourier domain. We also used 2695 MHz data from the Effelsberg
Galactic Plane Survey \citep{reic90}, whose angular resolution is
4\farcm3.

Characteristics of the new observations of DA\,495, made with the DRAO Synthesis
Telescope and the Effelsberg 100-m radio telescope, are presented in
Table~\ref{obspara}. 

\section{Results}

\subsection{DA\,495 in total intensity}

Figure~\ref{tp} presents total intensity images of DA\,495 at 408, 1420,
4850, and 10550~MHz as well as data at 2695~MHz from the Effelsberg
11\,cm survey \citep{reic90} and a 60$\mu$m image from the IRAS Galaxy
Atlas \citep[IGA,][]{cao97}, which is part of the CGPS database. The 1420~MHz
image is completely consistent with the earlier DRAO image published
by \citet{land83}, although the new image has lower noise by a factor
of $\sim3$. All the radio images show a diffuse source of full extent
about $25\arcmin$. The high-resolution images at 1420, 4850, and
10550~MHz show the central depression reported by \citet{land83}. Its
position is RA(2000) = $19^h 52^m 10.4^s$, DEC(2000) = $+29\degr
26\farcm2$, and it is now apparent that the pulsar (1WGA~J1952.2+2925)
is not at the center of the depression (as postulated by
\citet{land83}), but lies almost $2\arcmin$ east of it at RA(2000) =
$19^h 52^m 17.0^s$, DEC(2000) = $+29\degr 25\arcmin 53\arcsec$
\citep{arzo08}. Clearly the explanation of the hole in the nebula
cannot be quite as simple as a decline in emission from an aging
pulsar. At larger radii, the emission drops off smoothly towards the
outer edge without any sign of an outer shell at any frequency.

At 408~MHz DA\,495 is quite circular in appearance. However, an
extension of DA\,495 to the south-west is evident at 1420~MHz and
becomes increasingly prominent at higher frequencies. The IGA data
shown in Figure~\ref{tp} reveal an infrared source coincident with
this part of the radio source. The simplest interpretation is that an
\ion{H}{2} region is superimposed on the PWN, and in the analysis
we assume this to be the case.

\subsection{The integrated spectrum of DA\,495}

\begin{figure*}
  \centerline{\includegraphics[bb = 47 62 509 701,width=13cm,clip]{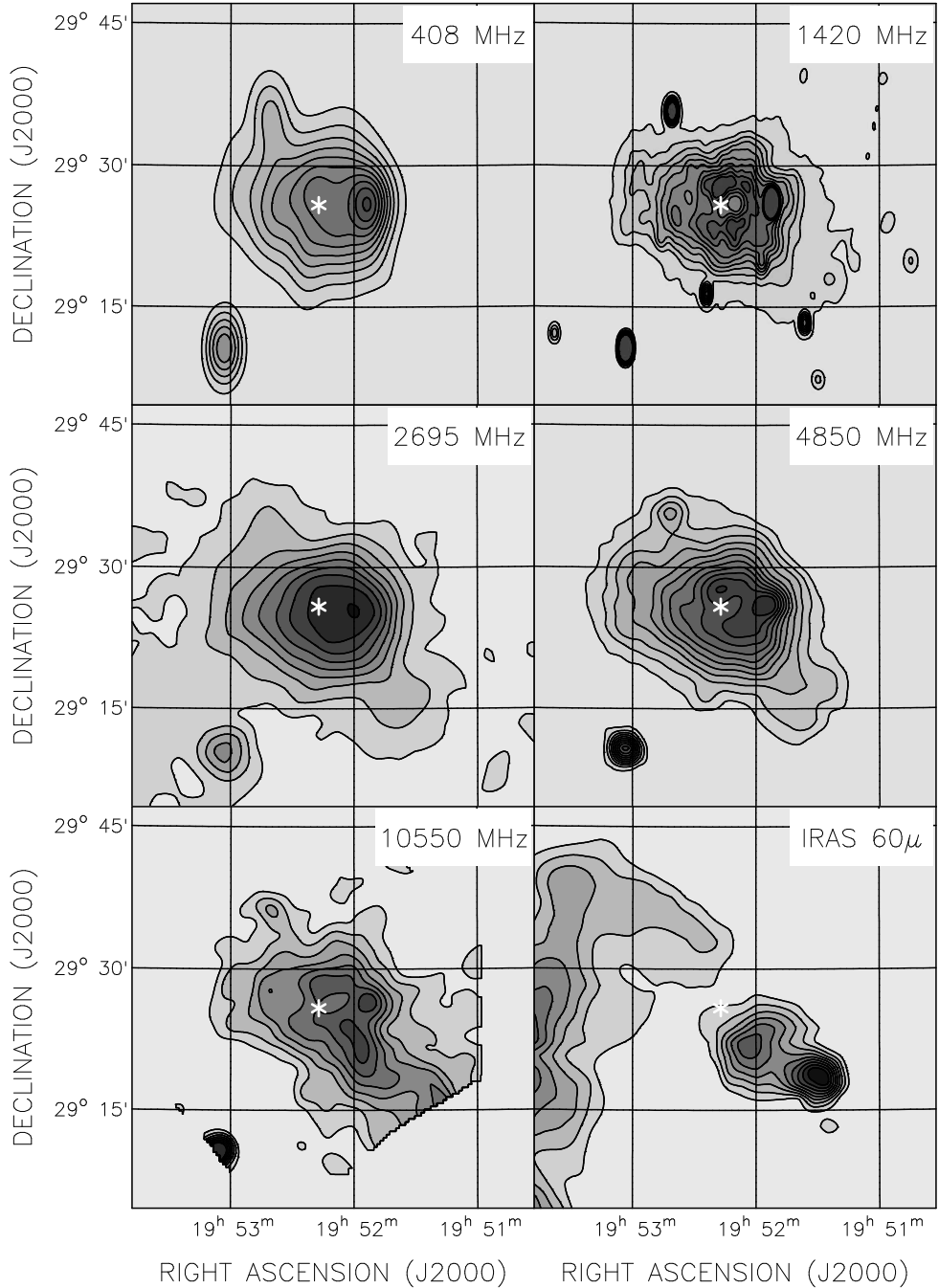}}
   \caption{Total intensity maps of DA\,495 at five radio
   frequencies, and the corresponding IRAS 60$\mu$m image.
   The position of the neutron star is shown in each panel by the white star.
   Contour levels are at: 10~K ($12\sigma$) to 100~K in steps of 10~K
   at 408~MHz (1~K = 8.5~mJy/beam); 0.5~K ($9\sigma$) to 5.0~K in steps of 
   0.5~K
   at 1420 MHz (1~K = 8.5~mJy/beam); 0.1~K ($5\sigma$) to 0.9~K in steps of 
   0.1~K
   at 2695 MHz (1~K = 400~mJy/beam); 5~mJy/beam ($20\sigma$) to 55~mJy/beam in
   steps of 5~mJy/beam at 4850 MHz (1~mJy/beam = 2.4~mK); 
   5~mJy/beam ($4\sigma$) to 
   40~mJy/beam in steps of 5~mJy/beam at 10550 MHz (1~mJy/beam = 0.5~mK);
   20 MJy/sr to 28 MJy/sr in steps of 1 MJy/sr at 60$\mu$m.}
   \label{tp}
\end{figure*}

We derived the integrated emission of DA\,495 from each of the radio
images in Figure~\ref{tp}, and identified a number of flux densities
from the literature. All flux densities which we believe to be
reliable are listed in Table~\ref{fluxes}. Flux density measurements
at low radio frequencies are not easy, especially for sources of low
surface-brightness; those that are available for DA\,495 need to be
discussed individually.

The area around DA\,495 is covered by the 34.5 MHz survey of
\citet{dwar90}, which has an angular resolution of $26\arcmin \times
44\arcmin$ at this declination. An apparently unresolved source of
flux density 35 Jy is detected, offset from the position of DA\,495 by
12\farcm5 in right ascension (to the west) and $5'$ in declination (to
the north). The offset may arise from compact sources included in the broad
beam (see below). The nearby SNR G65.1+0.6 \citep{land90} can be
identified in the 34.5 MHz image. Its expected flux density at 34.5
MHz, extrapolated from higher frequency measurements, is of the order
of 50 Jy, but its surface brightness is lower than that of DA\,495
because of its larger size ($1.5\degr \times 0.8\degr$). We consider
the 34.5 MHz data to be generally reliable in this area, and adopt an
upper limit of 35 Jy for possible emission from DA\,495 at 34.5
MHz. This is $\sim$7 times the rms sensitivity of 5 Jy quoted for the
survey.

The best low-frequency data come from the new VLA Low-Frequency
Sky Survey \citep[VLSS,][]{cohe07} at 74~MHz. We obtained an image of the area
surrounding DA\,495 from the 
VLSS data base\footnote{web address: http://lwa.nrl.navy.mil/VLSS/}.  The angular resolution is
$80\arcsec$, by far the best of any of our data points below 327~MHz,
and fully adequate to isolate the small-diameter sources in the
field. The survey is sensitive to structures up to $\sim$1$^{\circ}$
in size, larger than DA\,495. The survey noise is 100~mJy/beam, but
in the area around DA\,495 the noise is higher, about 150~mJy/beam,
because of the proximity to Cygnus~A. DA\,495 is not visible in the
74~MHz image, not even after smoothing to a resolution of
$3\arcmin$. To obtain an upper limit for the flux density at this
frequency, we convolved our 1420~MHz image to this resolution, and
added a scaled version to the 74~MHz image.  The scaled image was just
detectable when its integrated flux density was 14~Jy, and we adopted
this value as the upper limit of the 74~MHz flux density of DA\,495.
We tested this upper limit by adding our extrapolated 1420~MHz image
to areas around the position of DA\,495 as well. This convinced us
that we were not artificially lowering the upper limit by adding the 1420~MHz 
image to 74~MHz emission from DA\,495 that is just slightly below the 
detection threshhold.

Flux densities for DA\,495 at 83 and 111 MHz appear in the list
published by \citet{kova94}. These observations were made with fan
beams of extent a few degrees in the N--S direction. They should perhaps
also be regarded as upper limits because of the possibility of
confusion in the wide fan beams. We note that \citet{kova94}, while
listing flux densities for DA\,495, give only an
upper limit for the flux density of G65.1+0.6.

Taking the measured flux densities in Table~\ref{fluxes} at face value
we derive the integrated spectrum shown in the left panel of
Figure~\ref{intspec}, fitted reasonably well over the entire frequency
range by a spectral index of ${\alpha}={-0.59~\pm~0.1}$. However, the
compact sources in the field around DA\,495 and the superimposed
\ion{H}{2} region contaminate this spectrum. The compact sources are
probably extragalactic and should mostly have spectral indices
steeper than ${\alpha}={-0.59}$ \citep[e.g.][]{cond84}; they will 
consequently contribute
strongly to the low-frequency flux densities. The \ion{H}{2} region
will significantly contaminate the high-frequency flux densities
because its spectral index is flatter than that of the non-thermal
emission.  We now proceed to derive an integrated spectrum for DA\,495
corrected for these extraneous contributions.

We have obtained the flux densities of the three most prominent
compact sources in the field shown in Figure~\ref{tp} at frequencies
between 74~MHz and 10550~MHz, using our new data and values from
the literature. Flux densities of these sources are listed in
Table~\ref{sourcefd} and their positions and fitted spectra in
Table~\ref{sourceps}. The 74~MHz flux density for Source 1 is
taken from the VLSS point source catalog; the flux densities for
Sources 2 and 3 were derived by integration from the 74~MHz image.
Flux densities for DA\,495 after correction for compact sources are
shown in the fourth column of Table~\ref{fluxes}.

\begin{figure}
\centerline{\includegraphics[bb = 45 32 540 505,width=15cm,clip]{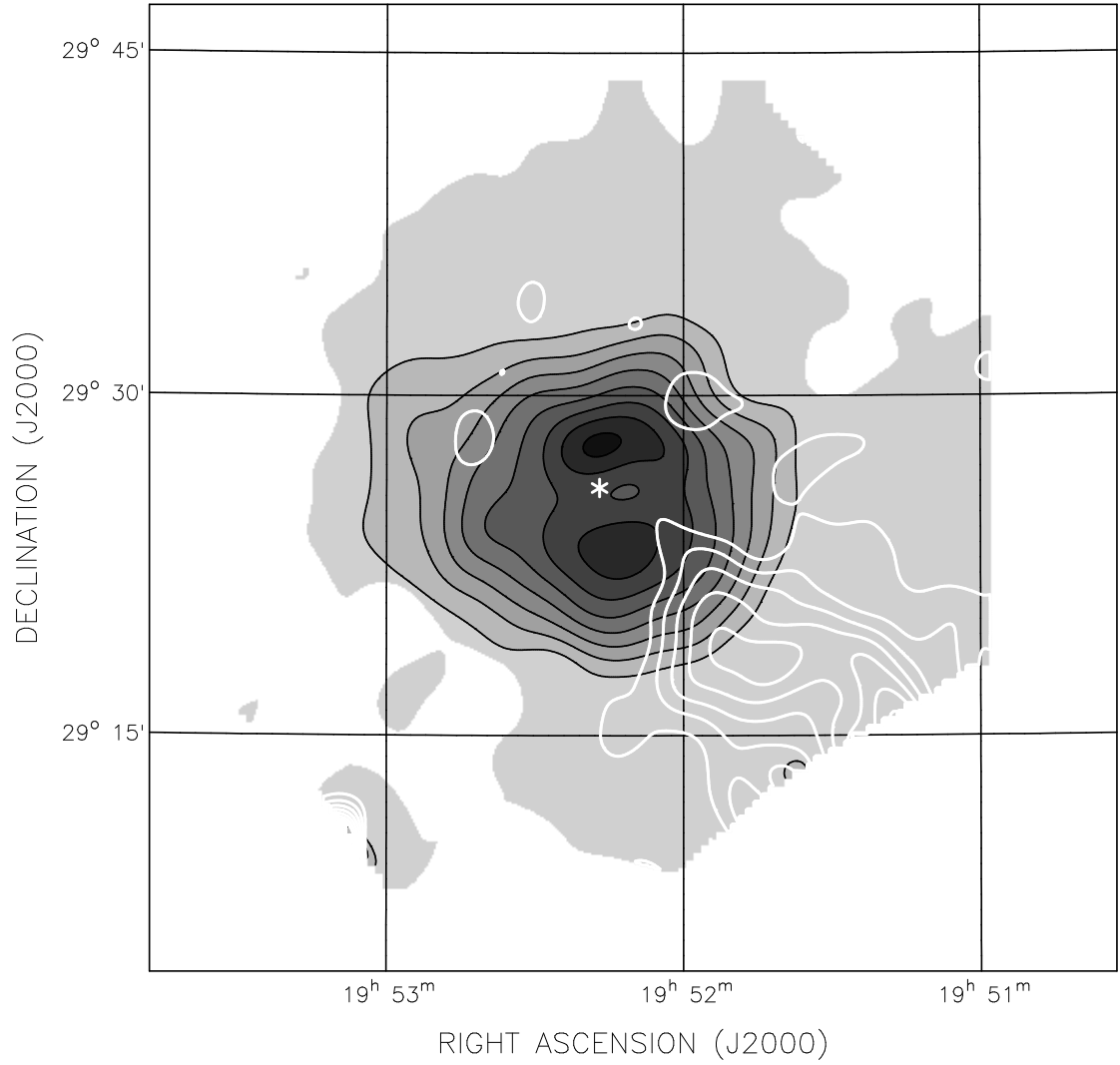}}
   \caption{Separation of the total power emission at 1420~MHz 
   into a non-thermal component (${\alpha}={-0.85}$ -- shown as black 
   contours and shading) and a thermal component (${\alpha}={-0.1}$ -- white
   contours). Black contours are from 30~mJy/beam ($10\sigma$) to 170~mJy/beam 
   in steps of
   20~mJy/beam. White contours are from, 10~mJy/beam ($3\sigma$) to 40~mJy/beam in 
   steps of 5~mJy/beam.
    The angular resolution is 2\farcm45 (see Section 3.2).}
   \label{nthth}
\end{figure}

\begin{figure*}
\centerline{\includegraphics[bb = 30 185 585 495,width=15cm,clip]{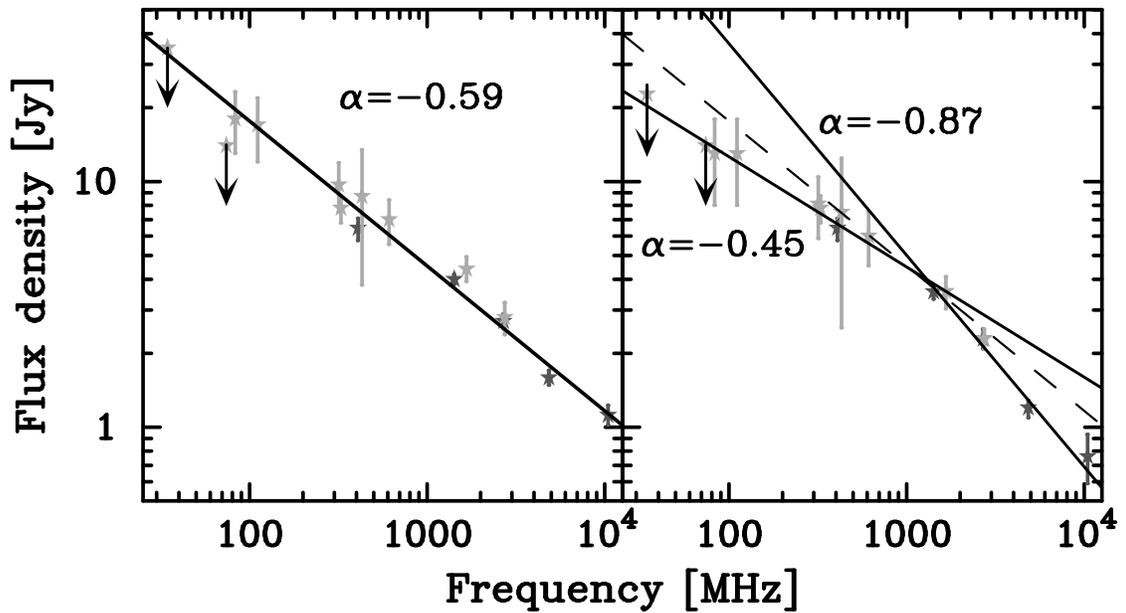}}
   \caption{Integrated radio continuum spectrum of DA\,495. Flux densities
   taken from the literature are displayed in gray and those
   determined in this publication in black. The left panel shows a
   spectrum constructed from the measured flux densities of
   Table~\ref{fluxes}. The indicated line, corresponding to
   ${\alpha}={-0.59}$, is a fit to all plotted points. The right panel
   shows the spectrum after allowance for the contribution of compact
   sources and the superimposed \ion{H}{2} region. The solid lines
   corresponding to ${\alpha}={-0.45}$ and ${\alpha}={-0.87}$ are fits
   to points below and above 1~GHz respectively (see Section 3.2). The dashed
   line corresponds to ${\alpha}={-0.59}$, the fit from the left panel.
   The lengths of the arrows denoting the upper limits are not intended to
   indicate error estimates.}
   \label{intspec}
\end{figure*}

We expect the \ion{H}{2} region to affect flux densities at high
frequencies, because its spectral index should be
${\alpha}\approx{-0.1}$ for optically thin thermal emission.  The
contribution from the \ion{H}{2} region was easily measured using our
images at 1420 MHz and 10550 MHz. After convolving to an angular
resolution of 2\farcm45, we plotted brightness temperature at 1420 MHz
against that at 10550 MHz (using images from which compact sources had
been subtracted) selecting only data from the north-east section of
the PWN which is unlikely to be contaminated by \ion{H}{2} region
emission. The comparison implied a spectral index of $-$0.85 for this
part of the PWN. Assuming that this is the spectral index of the
non-thermal emission between 1420 and 10550~MHz, and that the
\ion{H}{2} region has a spectral index of $-$0.1, we made a
point-by-point separation of emission into thermal and non-thermal
contributions. The result is shown in Figure~\ref{nthth}. The
non-thermal emission has strong circular symmetry around the central
depression, and closely resembles our 408~MHz image, where we expect
the contribution of the thermal component to be negligible. The
thermal component is confined to the south-west, and coincides closely
with the infrared source, as expected. Because of the similarity of the 
non-thermal map in Fig.~\ref{nthth} and the 408~MHz image in 
Fig.~\ref{tp}, the coincidence of the location of the thermal source 
in Fig.~\ref{nthth} and the infrared source in Fig.~\ref{tp}, and the 
successful separation, we can be confident that our assumption of a
constant spectral index for both the thermal and the non-thermal emission 
components is appropriate.

The contribution of the \ion{H}{2} region
to the flux density of DA\,495 is $440\pm 100$~mJy at 1420
MHz. We used this value with a spectral index of $-$0.1 to calculate
the contribution of emission from the \ion{H}{2} region at frequencies
of 1420 MHz and higher, and subtracted the calculated value to obtain
an integrated flux density for the emission from the PWN alone.  At
lower frequencies we did not take the emission from the \ion{H}{2}
region into account, since the contribution of the thermal emission
should become negligible and we cannot determine the frequency at
which the thermal emission becomes optically thick.  Flux densities
for DA\,495 after correction for the contribution of the \ion{H}{2}
region are shown in the fifth column of Table~\ref{fluxes}.

The right panel of Figure~\ref{intspec} shows the integrated spectrum
of DA\,495 after correction for both compact sources and the superimposed
\ion{H}{2} region. The corrected data strongly suggest a break in the
spectrum of DA\,495 at about 1 GHz.  Fitting spectra to frequency points
below 1 GHz yields a spectral index of
${\alpha}={-0.45~\pm~0.10}$. Fitting to flux densities above 1 GHz
yields ${\alpha}={-0.87~\pm~0.10}$. We consider the spectrum above
1~GHz to be well determined with a two-parameter fit to 6 well defined
flux density measurements. Below the break we have seven measured flux
densities two of which are upper limits. There are possible errors in the
spectral index below the break because of the poor angular resolution
in that frequency range. The quoted error ($\pm~0.10$) is the formal 
$1 \sigma$-error
in the fit, but there are probably larger systematic errors which we have
difficulty estimating; somewhat arbitrarily, we adopt a probable error
of 0.20.

These reservations notwithstanding, we believe that this analysis has
demonstrated convincingly that DA\,495 has a spectral break at
$1.3^{+0.3}_{-0.2}$~GHz of ${\Delta\alpha} = 0.42 \pm 0.22$. The spectral 
break may be
larger than 0.4, since the low-frequency flux densities are probably
upper limits, not definite measurements. Further, we have subtracted
the contributions of only three compact sources. Some contribution
from such sources will remain in our data, so that the plotted values
are again to some extent upper limits. The location of the 
break frequency, however, should be well defined. Its determination 
is dominated by the higher frequency fluxes of the low frequency part 
of the spectrum. These values are well constrained, while some of the low
frequency fluxes are upper limits. These affect only the slope of
the spectrum below the break.

\subsection{Polarization images and rotation measure}

\begin{figure*}
\centerline{\includegraphics[bb = 45 60 511 505,width=15cm,clip]{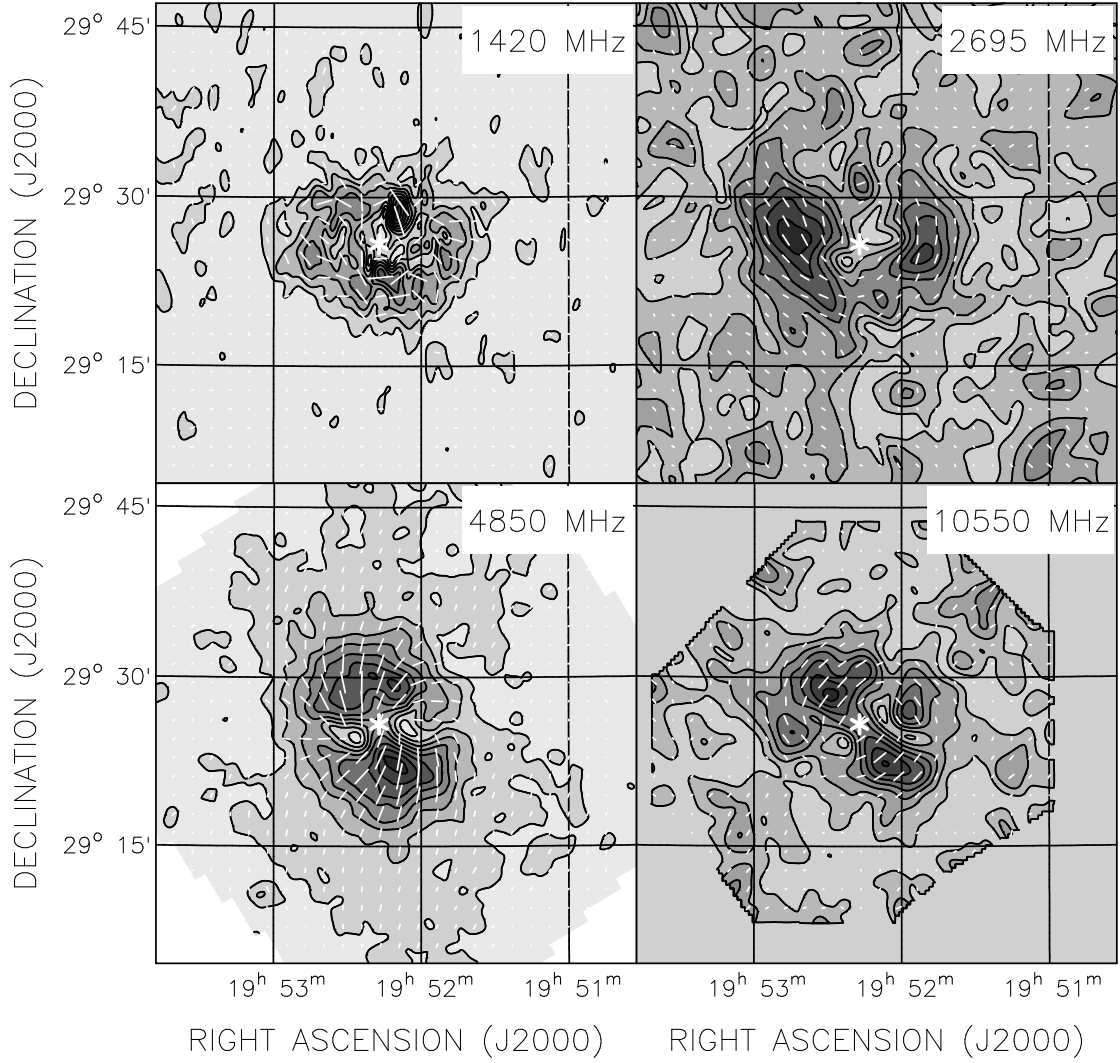}}
   \caption{Polarized intensity of the emission from DA\,495 and its 
   surroundings at four frequencies, shown by contours and grayscale.
   The orientation of the E-field is shown by the superimposed vectors.
   The location of the pulsar is indicated in each panel 
   by the star.
   Contour levels are at:
   1.5~mJy/beam ($4\sigma$) to 12~mJy/beam at 1420~MHz; 
   6~mJy/beam ($2\sigma$) to 42~mJy/beam
   in steps of 6~mJy/beam at 2695~MHz; 
   1.2~mJy/beam ($6\sigma$) to 9.6~mJy/beam
   in steps of 1.2~mJy/beam at 4850~MHz; 
   0.8~mJy/beam ($2\sigma$) to 5.6~mJy/beam
   in steps of 0.8~mJy/beam at 10550 MHz.}
   \label{pi}
\end{figure*}

Figure~\ref{pi} shows our polarization images for the area around
DA\,495, together with data from the Effelsberg 2695 MHz survey
\citep{dunc99}. At all frequencies the polarized emission is confined
to a roughly circular region about $15\arcmin$ in diameter. No
polarized emission is seen corresponding to the south-west extension,
supporting the identification of this as thermal emission. No
depolarization effects can be associated with this thermal emission
either, and the \ion{H}{2} region which generates it is likely more 
distant than DA\,495. 

Integrated over the entire source, the fractional polarization at the
higher frequencies is about 25\% (Table~\ref{intpol}). Even at
1420~MHz the integrated polarization has dropped only to 12\%, still
half the value at the higher frequencies. At 1420~MHz a RM variation
of only 36~rad/m$^2$ is required to rotate the polarization angle by
$90\degr$, and superposition of two signals whose polarization angles
differ by $90\degr$ produces a net signal with no apparent polarization. 
(At our other frequencies RM variations of about
130~rad/m$^2$ at 2695~MHz, 400~rad/m$^2$ at 4850~MHz, and
1900~rad/m$^2$ at 10550~MHz are required to generate a Faraday
rotation of $90\degr$.) This implies that there cannot be any rapid
changes in rotation measure on scales smaller than the 1420~MHz beam,
because otherwise the 1420~MHz emission would be beam depolarized.

At the two highest frequencies the polarized intensity has a
remarkable bipolar distribution, seen particularly well in the
4850~MHz image (Fig.~\ref{pi}).  Inspection shows that the bipolar
structure is centered on the pulsar, immediately suggesting that the
polarization structure is determined by the geometry of the magnetic
field, itself tied to the pulsar. We see two prominent lobes of
polarized emission, north-east and south-west of the pulsar. At the
three high frequencies, from 10550 to 2695~MHz, the location of these
lobes seems to rotate systematically clockwise, in the sense opposite
to the rotation of polarization angle that arises from the positive RM
(see below). Symmetrically on either side of the central ridge at 4850
and 10550~MHz there are small regions of low polarized intensity.
The total intensity remains high at these locations, and we
interpret the low fractional polarization as an indication that
polarized emission generated at deeper layers within the source is
superimposed on emission generated in closer layers with a
significantly different polarization angle; vector averaging reduces
the net polarization. This is largely an emission effect rather than a
Faraday rotation effect because it is seen at high radio frequencies
where Faraday rotation is low.

With linear polarization observations at four frequencies we should be
able to investigate the structure of RM across the source.  To this
end, all observations were convolved to the lowest resolution,
$4\farcm3$, that of the 2695~MHz measurement. However, it proved
impossible to calculate a RM map: at many points the calculation
failed to fit a slope to the four measured polarization angles within
acceptable error limits. The probable cause is ``Faraday thickness''
at the lower frequencies, an effect analogous to optical thickness.
Depth depolarization - also known as differential Faraday rotation
\citep{burn66,soko98} -
causes the emission from the
deeper levels of the source to be totally depolarized at low
frequencies, so that the polarized emission we observe comes only from
the nearer layers of the source.  Faraday thickness seems to affect
the polarization images at 2695~MHz and 1420~MHz. 

\begin{figure}
\centerline{\includegraphics[bb = 30 60 555 560,width=15cm,clip]{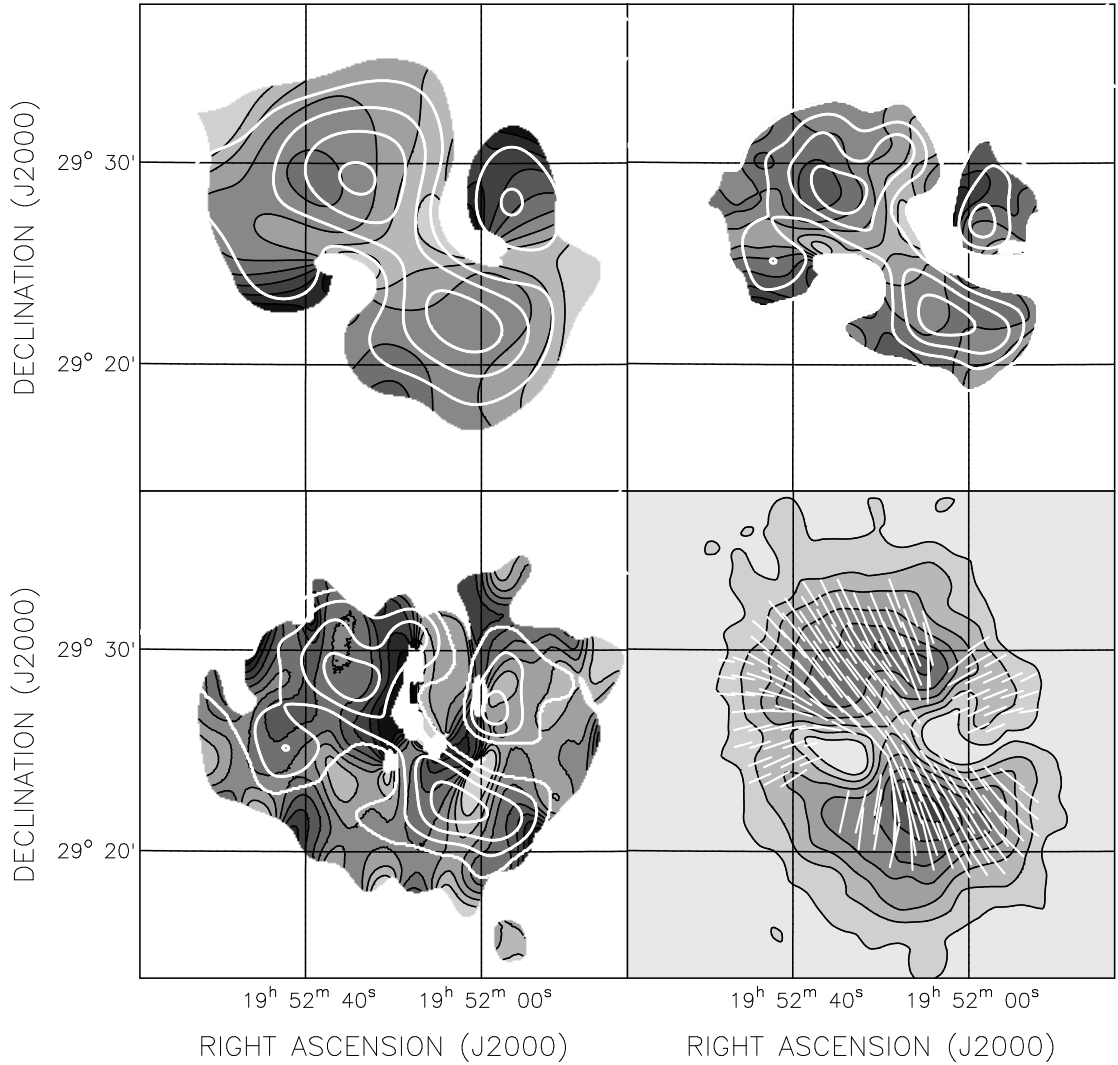}}
   \caption{Upper panels: Maps of rotation measure between 4850~MHz
   and 10550~MHz, at the resolution of the 2695~MHz map, $4\farcm3$
   (left) and at a resolution of $2\farcm45$ (right). Black contours
   represent rotation measure, with values 140, 180, 220, 260, 300,
   340, 380, and 420~rad/m$^2$. White contours indicate polarized
   intensity at 10550~MHz at the appropriate resolution. Lower panels:
   Left: Map of rotation measure between the four frequency bands of
   the CGPS at $2\farcm45$ (left). Contour levels are the same as in
   the other RM maps. Right: Map of polarized intensity at 4850~MHz
   shown by contours and grayscale, with levels 1.2 mJy/beam to 9.6
   mJy/beam in steps of 1.2 mJy/beam. Overlaid white vectors indicate
   the intrinsic magnetic field directions calculated using the RM map
   at the top right.}
   \label{rmpib}
\end{figure}

Confining our attention to 4850 and 10550~MHz, we calculated a map of
RM at a resolution of 4\farcm3, displayed in Figure~\ref{rmpib}.  RM
varies from 200 to 250~rad/m$^2$ on the two major emission lobes,
north-east and south-west of the pulsar, to as much as 400~rad/m$^2$
in the structures to the east and west.  To identify the Faraday thin
and Faraday thick areas we calculated the expected polarization angles
at 2695~MHz by extrapolating from the observed angles at 4850 and
10550~MHz and compared the results with values actually observed at
the lower frequency. The polarization angle in the south-western lobe
is as predicted, while the angle in the north-eastern lobe is about
$50^{\circ}$ from the predicted value. Observed polarization angle in
two patches lying approximately $4'$ east and west of the pulsar
(these patches are prominent in the 10550~MHz image of
Figure~\ref{pi}) is about $90^{\circ}$ from the predicted
value. Faraday thickness is clearly significant in this object.
We also calculated a RM map between 4850 and 10550~MHz at the
higher resolution of $2\farcm45$ to examine RM fluctuations on smaller
scales. The RM maps at the lower and higher resolution look virtually
identical, confirming our impression that there are no strong RM
fluctuations on small scales. We also calculated intrinsic
polarization angles at the highest resolution and derived a map of the
magnetic field (projected to the plane of the sky). This is also shown
in Fig.~\ref{rmpib}.

Even though DA\,495 is Faraday thick at frequencies of 2695~MHz and
below, we calculated a RM map from the four bands of the CGPS 1420~MHz
polarization data (Fig.~\ref{rmpib}). Surprisingly, these values of
RM, which probe only the nearer layers of the source, are of the same
order ($\sim 200$~rad/m$^2$) as the values derived between 4850 and
10550~MHz where the entire line of sight through the PWN is
contributing. If RM was positive right through the source the values
at high frequencies should be higher than the values near
1420~MHz. This indicates that there must be regions of negative RM in
the deeper layers whose effects are canceled by positive rotation
nearer the front surface. A possible field configuration that
could produce this is a dipole field with the dipole axis pointing
away from us. Almost the entire front part of the nebula would contain
magnetic field lines that point towards us, explaining why we observe
only positive rotation measure at 1420~Mhz.  However, deeper inside
the nebula we would find magnetic field lines pointing away from us,
and the Faraday rotation produced by that part of the field would
partially cancel rotation produced in the near part. The exact RM
configuration seen would depend on how deep we can look inside the
nebula at 1420~MHz or other frequencies.

\section{Discussion}

\subsection{Is DA\,495 a PWN or a Shell SNR?}

In the introduction we reviewed the arguments for the classification
of DA\,495 as a PWN. In brief, the emissivity is centrally
concentrated and drops smoothly to zero at the outer edge without any
sign of a steep shell, indicating a central energy source. This
conclusion is corroborated by X-ray observations \citep{arzo04,arzo08} 
that show a central compact object surrounded by a
non-thermal nebula. \citet{velu89} suggest that DA\,495 could be a 
shell remnant with an unusually thick shell. While this is a conceivable 
interpretation, there is no evidence of a steep outer edge to the shell
and the central compact source discovered in X-ray is not located
inside the radio hole, but about 2\,$\arcmin$ away from it suggesting
that the radio hole is not in the centre of the nebula.

This paper provides further evidence. The structure in polarized
intensity is bipolar with mostly radial B-vectors at the outer edge,
while for a shell-type remnant we would expect a shell structure in
polarized emission with tangential B-vectors. A young shell-type SNR
would have radial B-vectors, but integrated polarization would be very
low and a young remnant would have bright X-ray emission coming from
the expanding shell. The break in the integrated spectrum of DA\,495
also provides strong evidence that the object is a PWN. Every
known example of such a nebula exhibits such a break, while
virtually all shell remnants have spectra characterized by a single
power law. While spectral breaks have been claimed for some shell
remnants, these claims are gradually melting away in the face of new
data with improved sensitivity to extended emission.  The only SNR
where such a break has been convincingly established is S\,147
(G180.0$-$1.7) which has a break in its integrated spectrum at
$\sim$1.5~GHz \citep{furs86,xiao08}. S\,147 is a very old SNR, probably well
into the isothermal stage of its evolution; it bears no resemblance to
DA\,495.

The only tenable conclusion is that DA\,495 is a pulsar wind nebula, and we
will proceed on that basis. 

\subsection{Distance, Size, and Emission Structure}

A distance to DA\,495 was measured by \citet{koth04} on the basis of
absorption by \ion{H}{1} of the polarized emission from the PWN.
The distance obtained, $d \approx  1.5$\,kpc, was derived kinematically using
a flat rotation model for the Milky Way with a Galactocentric
radius of R$_\sun = 8.5$\,kpc for the Sun and its velocity
v$_\sun = 220$\,km/s around the Galactic center. The latest measurements 
of the Sun's
galactocentric distance give R$_\sun \approx 7.6$\,kpc
which is used in the new distance determination method of
\citet[and reference therein]{fost05}.  This method uses a model for the
spatial density and velocity field traced by the distribution of
\ion{H}{1} in the disk of the Galaxy to derive a distance--velocity
relation that can be used to calculate a distance to any object with a
known systemic velocity.  The \ion{H}{1} absorption measurements indicate
for DA\,495 a systemic velocity of about $+12$\,km/s. Two distances, about
1 and 5\,kpc, correspond to this velocity, one closer than the tangent
point and the other beyond it.  Since \citet{koth04} found no
absorption of the PWN at the tangent point and DA\,495 exhibits low
foreground absorption in the X-ray observations \citep{arzo04,arzo08}
we adopt the closer
distance as the more likely result.  We proceed to use a distance
of $1.0\pm 0.4$\,kpc to determine physical characteristics of the PWN.
The error contains the uncertainty in the fit and a velocity dispersion of
4\,km\,s$^{-1}$.

To determine a reliable value for the radius we fitted a
three-dimensional model of emissivity to the total-intensity
observations at 1420~MHz. Our method is similar to that used by
\citet{land83}, a method developed by \citet{hill67}. While
\citet{land83} centered their calculation on the depression near
the middle of the nebula, we centered our new calculation on the
position of the pulsar, as measured by \citet{arzo08}. In the center 
there is an emissivity depression
with a radius of about $2\arcmin$, and so a physical radius of
0.6\,pc.  Outside, the emissivity is nearly constant to a
radius of $7\farcm5$, where it drops abruptly, almost to zero. This
translates to an outer radius of 2.2\,pc. \citet{land83} found that
the emissivity increased from the outer edge towards the
centre. This difference can be explained by the different location
assumed for the centre and the contamination by the \ion{H}{2}
region (removed from our calculation but included in that of
\citet{land83}).

\citet{land83} assumed that
the prominent depression marked the center of the 
PWN, but we
know now that the pulsar lies at one edge of that depression
\citep{arzo04,arzo08}, not at its center. One might suppose that the pulsar
has moved since it was formed, but we see from Figs.\,\ref{pi} and
\ref{rmpib} that the structure in polarized intensity, and so the
magnetic field structure, is centered on the pulsar. Since the radio
appearance of the nebula is dominated by particles injected early in
the life of the pulsar (see Section 4.3), it appears that the pulsar
has {\it{not}} moved significantly, and we must find another
explanation.  

\begin{figure}
  \centerline{\includegraphics[bb = 45 30 540 505,width=15cm,clip]{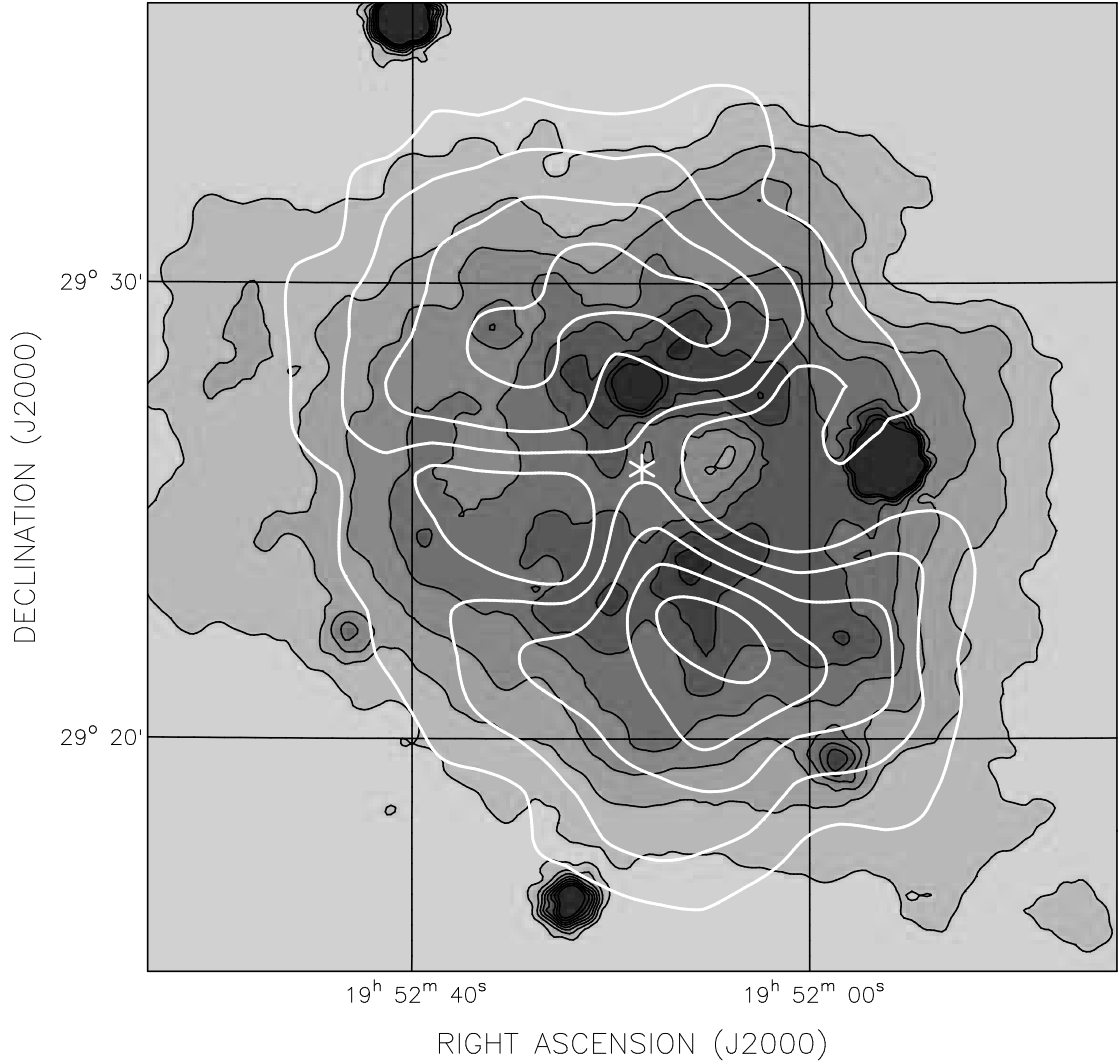}}
   \caption{Map of total intensity at 1420~MHz made by merging
   the CGPS 1420~MHz map with a map taken from the NVSS (see Section 4.2).
   The resolution is about $45\arcsec$. Contour levels are at
   0.8 ($8\sigma$), 1.6, 2.4, 3.2, 4.0, 4.8, and 5.6\,K. Overlaid white 
   contours indicate the polarized intensity at 4850\,MHz with 
   levels at 7, 10, 13, 16, and 19\,mJy/beam.}
   \label{tppi}
\end{figure}

In order to achieve higher angular resolution at
1420~MHz, we merged the 1420 MHz map from Fig.\,\ref{tp} with data
from the NRAO VLA Sky Survey \citep[NVSS;][]{cond98}. The NVSS has a
resolution of about $45\arcsec$.  The resulting map is shown in
Fig.\,\ref{tppi}.
Inspection of Fig.\,\ref{tppi} shows evidence of a
second less obvious depression in total intensity, at
RA(2000)\,=\,19$^h$ 52$^m$ 33$^s$, DEC(2000)\,=\,29$^{\circ}$
26$'$. This eastern depression is further from the pulsar than the
more obvious, and deeper, hole to the west, and the nebula appears to extend to
slightly larger radius beyond the eastern hole. Nevertheless, we
regard the eastern depression as a significant feature.  In order to
produce these two emissivity depressions, there must be a region around the
pulsar which is relatively free of magnetic field and/or particles.
This brings to mind the Crab Nebula's ``synchrotron
bays''. \citet{fese92} ascribe them to an equatorial torus of material
ejected by the progenitor star before it exploded. They suggest that
the magnetic field of the nebula is wrapped around this torus
preventing relativistic particles from entering it. A saddle-shaped
depression in total power emission that connects the two holes in
DA\,495 (see Fig.~\ref{tppi}) supports a torus as an explanation for
these features. In the Crab Nebula the synchrotron bays are located at
the edge of the emission region, but DA\,495 is probably much older
and the nebula has had time to flow around the torus. The distance
between the two holes is about $5\arcmin$ translating to 1.45~pc at a
distance of 1~kpc. This indicates an average expansion velocity of
about 70~$\frac{10000\,yr}{t}$~km/s, where $t$ is the age of DA\,495.
For an age of 10000~yr this expansion velocity would be about 10~\% of
that of the Crab Nebula's synchrotron bays, which are expanding at
about 900 and 600 km/s for the western and eastern bay respectively
\citep{fese92}. We cannot completely rule out that movement
of the pulsar is responsible for the off-center location of the
pulsar between the two holes, but a difference in expansion velocity
between the two sides, similar to that observed in the Crab Nebula,
could explain this asymmetry.

\subsection{The Nature of the Spectral Break}

To investigate the nature of the break in the radio spectrum of
DA\,495 we first estimate the nebula's age by comparing
its energetics with the properties of other PWNe of known age.
We assume for the moment that the pulsar has a symmetrical pulsar wind
and that equipartition between particle
and magnetic field energies has been established. We follow the analysis
made by \citet{koth06a} of the PWN G106.65+2.96, although, unlike that case,
we have no information on the characteristic age of the DA\,495 pulsar,
which has not been detected. Nevertheless, we can get an estimate of
the current energy loss rate of the pulsar from \citet{arzo08}:
$\dot{E}_{\rm DA\,495} \approx 10^{35}$~erg/s. 
We now confine
our attention to pulsars in historical SNRs for which we know the
exact age, the Crab Nebula (SN 1054), 3C~58 (SN 1181), and G11.2$-$0.3
(SN 386) \citep{gree04}. We also add three pulsar wind nebulae for which
\citet{koth06a} have calculated the real ages from the location of
the cooling breaks in their
radio spectra. We extrapolate their evolution to
$\dot{E}_{\rm DA\,495}$ to determine possible properties of the pulsar
in DA\,495.  

The evolution of the energy loss rate, $\dot{E}$, of a
pulsar is defined by \citep[e.g.][]{paci73}:  
\begin{equation} 
\dot{E}(t) = \frac{\dot{E_0}}{(1+\frac{t}{\tau_0})^\beta}, 
\end{equation} 
where
$\beta$ is related to the nature of the braking torque {($\beta =
\frac{n + 1}{n-1}$, and $n$ is the braking index of the pulsar)}, 
$t$ is the real
age of the pulsar, $\dot{E_0}$ is its energy loss rate
at the time it was born, and $\tau_0$ is its intrinsic characteristic age
($\tau_0 =
\frac{P_0}{2\dot{P}_0}$, where $P_0$ is the period of the pulsar at
the time it was born). Using the information about the current characteristic 
age $\tau$ and
$\dot{E}$ for the six pulsars taken from the ATNF Pulsar
Catalogue \citep{manc05}\footnote{web address:
http://www.atnf.csiro.au/research/pulsar/psrcat/} and their real age
we can derive their initial parameters, $\tau_0$ and $\dot{E}_0$. For
the Crab Pulsar and the pulsar in Vela~X the braking indices are
known, yielding a $\beta$ of 2.3 \citep{lyne88} and 6.0
\citep{lyne96}, respectively. For the other four pulsars we assume a
pure dipole field, implying a $\beta$ of 2. For Vela~X the age depends
on the assumed value of $\beta$, since we calculate it from the
spectral break frequency. The resulting intrinsic pulsar properties
are listed in Table~\ref{pulsars}. To indicate how much our
calculations depend on this factor we used different values of $\beta$
as indicated in Table~\ref{pulsars}.  Using equation 1 and taking the
energy loss rate of the pulsar in DA\,495 as $\dot{E}_{\rm DA\,495}
\approx 10^{35}$~erg/s, we can calculate the current age, $t_{\rm
DA\,495}$, of the nebula and the current characteristic age,
$\tau_{\rm DA\,495}$, of its pulsar using the initial values for each
of the six pulsars.  The results are given in
Table~\ref{DA495_pulsar}. For the Crab and the Vela~X pulsars we
extrapolated the energy loss rate with two different $\beta$ values as
indicated in Table~\ref{DA495_pulsar}. Whatever basis we adopt for the
calculation, the age of DA\,495 exceeds 20,000 yr, which makes this
nebula an old object. This derived age is valid regardless of the
mechanism responsible for the break in the radio spectrum. 
All known PWNe detected at radio wavelengths contain energetic pulsars, so 
the low energy loss rate of the DA\,495 pulsar is in all likelihood the 
result of
age, not of intrinsically low energy.

We will now proceed by examining the different possibilites for a spectral
break in the synchrotron spectrum of a pulsar wind nebula.
Virtually all PWNe exhibit breaks in their spectra, at
frequencies ranging from X-rays to a few GHz.  Table~\ref{break_freq}
lists the break frequencies that have been identified to date. The
spectral break in DA\,495 occurs at 1.3\,GHz, significantly lower than
for any other pulsar wind nebula. A spectral break in a PWN can 
arise in two
ways: it may reflect a break in the injected electron spectrum, and so
be a mostly pulsar-dependent effect, or it may 
arise from synchrotron
cooling, in which case the break frequency depends on the magnetic
field strength and the age of the relativistic electrons in the
nebula. Of course,
both mechanisms should be at work in all PWNe. The suspected nature of
each spectral break is indicated in Table~\ref{break_freq}.

We first consider the possibility that the break in the radio
spectrum results from a break in the injected electron spectrum.
Such breaks in the spectra of PWNe are known (e.g. in the Crab
Nebula and 3C~58), but they are imperfectly understood.  Since the
electron population in the nebula is dominated by early injection, the
break must have been created early in the life of the pulsar.
The energy at which the break occurs depends on conditions in the
area in which electrons are accelerated. Its location in the electron
spectrum remains relatively fixed throughout the life of the nebula. In 
contrast, a break due to
cooling, starts at high energy and moves with time towards lower
energies \citep[e.g.][]{weil80} ultimately perhaps passing by the
break produced by the injected spectrum. If the magnetic field decays
with time, through expansion or other causes, both kinds of break will
migrate towards lower frequencies in the radio spectrum. Their 
relative positions will not be affected.

A likely location for an injected break is at the energy where the
electron acceleration timescale equals the synchrotron loss time,
which is in essence a kind of cooling break. The difference between
this situation and the normal synchrotron cooling break is that the
number of high energy electrons is never high at any stage. Thus, the
synchrotron spectrum should steepen by much more than the 0.5 typical
of normal synchrotron cooling, which is actually the case for the
intrinsic breaks identified in other PWNe. For DA\,495 we found that
the spectrum steepens by about 0.4.

There are two other arguments that make an injected break an
unlikely interpretation for the break in the radio spectrum of
DA\,495.  First, the X-ray synchrotron spectrum has a spectral index
of $-0.63 \pm 0.27$
\citep[][calculated from the photon index $\Gamma$:
$\alpha = - (\Gamma - 1)$]{arzo08}, flatter than the synchrotron radio spectrum above the
break ($-$0.87).  This is impossible if the radio break arises from an
injected break. The only mechanism that can affect the synchrotron
spectrum beyond an intrinsic break is cooling or some other energy
loss effect, which can only steepen the spectrum but cannot flatten
it. A new round of particle acceleration late in the life of the PWN,
an unlikely event, would affect low-energy and high-energy electrons
alike. Second, there is no injected break frequency known that is even
remotely as low as the 1.3~GHz we have found in DA\,495.

We now consider the possibility that the break at 1.3~GHz is caused by 
synchrotron cooling. We will assume that the magnetic field 
throughout DA\,495 is relatively constant or at least that there is
a dominating magnetic field creating the cooling break. There are several 
lines of
evidence supporting this assumption. First, our calculation of the
synchrotron emissivity from the total-intensity images shows that
emissivity is nearly constant in the majority of the nebula (Section
4.2). Second, the break in the total spectrum is very sharp and there
is no significant variation of spectral index across the source
(DA\,495 has the same appearance at all radio frequencies, see also
section 3.2: the separation into thermal and non-thermal components).  
Whatever
the cause of the spectral break, it is the magnetic field that
translates a break in the electron energy spectrum to a break in the
emission spectrum.  If the magnetic field varied significantly, the break
would be smeared across a wide range of frequency; the break frequency
has a very strong dependence on field strength (see Equation (2)
below).

\citet{chev00} found that the cooling frequency, $\nu_c$, corresponding 
to $\gamma$ (the Lorentz factor of the electron) at which 
the electrons are able to radiate their energy over the age of a 
PWN, is defined by
\begin{equation}
   \nu_c [{\rm GHz}] = 1.187\cdot B^{-3} [{\rm G}] \cdot t^{-2}
[{\rm yr}],
\end{equation}
where $B$ is the magnetic field strength inside the synchrotron nebula. Using
the break frequency of $\nu_c
\approx 1.3$~GHz and the various interpolated ages, $t_{\rm DA\,495}$, we can
calculate $B_{\rm{req}}$, the required magnetic field inside the
nebula (Table~\ref{DA495_pulsar}). Integrating equation (1) over the age
of the nebula gives the total energy inside it and, assuming
equipartition and a circular structure, we can calculate the magnetic
field, $B_{\rm{max}}$. We denote this field value $B_{\rm{max}}$
because we have not accounted for any energy loss in the course of the
evolution of the PWN. Comparing the results based on the six pulsars
listed in Table~\ref{DA495_pulsar}, we see that it is only in 
the calculation based on the Crab pulsar and
the pulsar in Vela~X that the maximum magnetic 
field exceeds the magnetic field required for the spectral break.
If our assumption is valid, then the pulsar in DA\,495 
must have 
been very energetic with a high energy loss rate at the time of formation,
like the Crab pulsar or the pulsar in Vela~X. Given the age of the
Vela supernova remnant and the fact that $B_{\rm{max}}$ is 
significantly higher than
$B_{\rm{req}}$ for any $\beta$ it is more reasonable to assume that
the pulsar in DA\,495 is an older version of the Vela pulsar. If we shrank
the Vela~X nebula with its current energy content to the size of
DA\,495, its cooling break at 100~GHz would decrease to about 2.5~GHz
due to the resulting increase in the magnetic field strength.
A few thousand years later the break frequency would
reach 1.3~GHz.

Even though we cannot completely rule out the possibility that the
break in the radio spectrum of DA\,495 at 1.3~GHz is caused by a break
in the injected electron spectrum, all the available evidence
indicates that synchrotron cooling is the most likely explanation, and
that DA\,495 is probably an aging Vela~X nebula. In this case DA\,495
is about 20,000~yr old and contains a strong magnetic field of
about 1.3~mG. This can easily account for the high rotation
measure we have observed.

\subsection{The magnetic field structure inside DA\,495}

The structure of the PWN in total intensity is very regular,
especially once the confusing \ion{H}{2} region is removed (see
Figure~\ref{nthth}). However, in polarized intensity the appearance is
very different, and at the high frequencies DA\,495 is very
structured.  Percentage polarization, even at the high frequencies,
varies over a wide range across the source. The challenge faced in
this section is to deduce something about the structure of the
magnetic field inside DA\,495 and to explain as many of the observed
characteristics as possible.  While we cannot uniquely determine
the field configuration we show that one relatively simple and
plausible model, a dipole field with a superimposed toroidal
component, is consistent with the observations.

\subsubsection{The foreground Faraday rotation}

\begin{figure*}
\centerline{\includegraphics[bb = 60 50 565 760,angle=-90,width=16cm,clip]{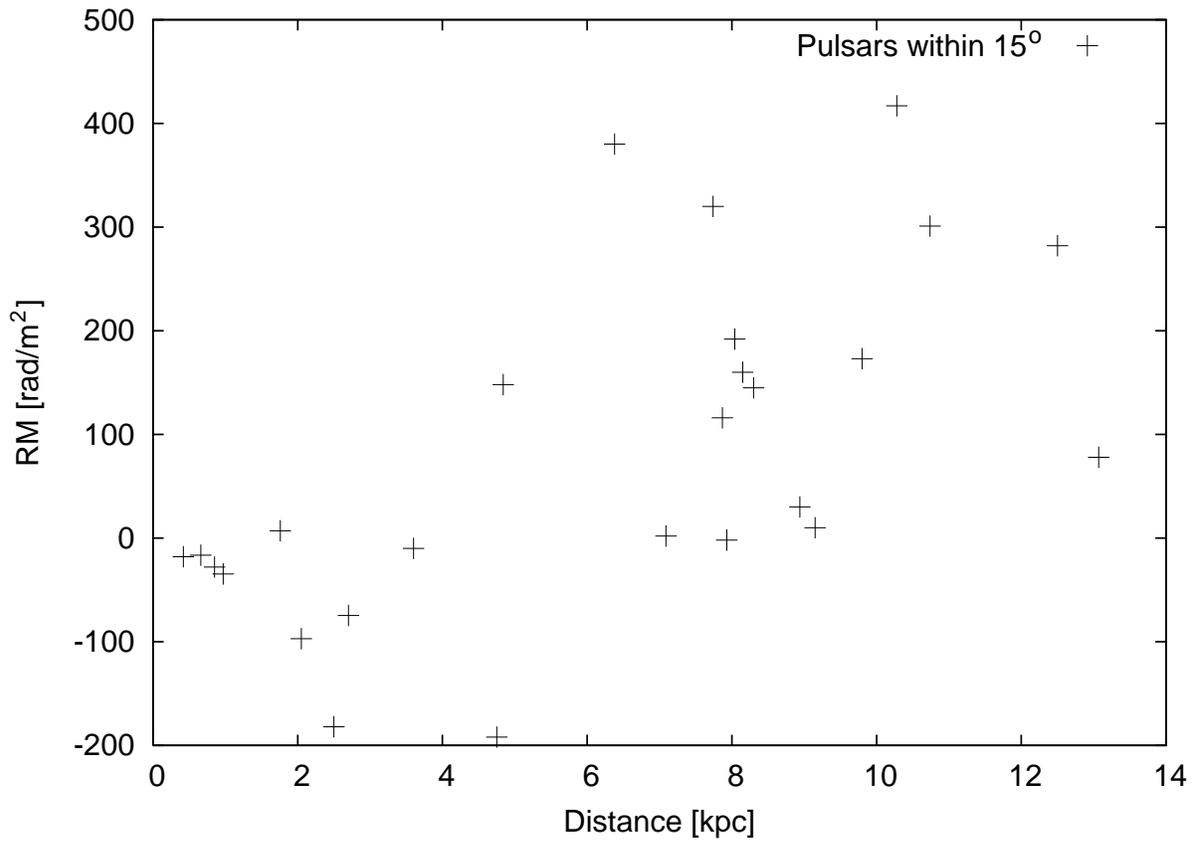}}
   \caption{Foreground rotation measure for all pulsars within $15\degr$ 
   of DA\,495 as a function of distance from us in kpc. 
   See Section 4.4.1 for source of data.}
   \label{pulsarrm}
\end{figure*}

In Figure~\ref{pulsarrm} we have plotted the foreground RM observed to
all pulsars within $15\degr$ of DA\,495 as a function of distance 
from us in kpc.  The pulsar data were taken from the ATNF Pulsar
Catalog \citep{manc05}\footnote{web address:
http://www.atnf.csiro.au/research/pulsar/psrcat/}.  All pulsars within
about 5~kpc have negative RM values, except for one for which
${\rm{RM}}={+7{\rm{~rad/m}}^2}$; we will disregard this exception
because it lies at a Galactic latitude of $+14\degr$. The
negative RM values are consistent with the known local magnetic field
orientation, away from the Sun. Pulsars more distant than 5~kpc have
positive RMs, as expected from the well-known field reversal between
the local and Sagittarius spiral arms 
\citep[e.g.][]{sima79,lyne89}.  Averaged over the nearby pulsars the
line-of-sight component of the magnetic field is around $-$2.5~$\mu$G.
Using the models of \citet{tayl93} and \citet{cord02} for the
distribution of free electrons in the Galaxy, and the distance to
DA\,495 of 1.0~kpc we determine the average foreground dispersion
measure to be ${\rm DM} \approx 15$~cm$^{-3}$~pc.  With an average
foreground magnetic field of $-$2.5~$\mu$G parallel to the line of
sight this produces a foreground RM of $-$30~rad/m$^2$. The positive
RM values determined from the emission of DA\,495 must arise from
Faraday rotation within the PWN itself or occur in a discrete
foreground magneto-ionic object which does not itself produce enough
emission to be detected in our observations. Since we find that almost
all of the polarized emission from DA\,495 is Faraday thick at
2695~MHz, most of the positive RM must originate in DA\,495, because
Faraday thickness can only be produced if the rotation is generated in
the same volume elements as the polarized emission itself. A
foreground object can merely rotate or beam depolarize the entire
emission coming from behind it.

\subsubsection{Magnetic field configurations inside DA\,495}

The magnetic field distribution seen in Figure~\ref{rmpib} is the
projection onto the plane of the sky of the three-dimensional field
within DA\,495, as deduced from our observations at 4850 and 10550~MHz
on the assumption that the source is Faraday thin at these
frequencies. The first impression is of a highly regular 
elongated magnetic field configuration with field lines following a
central bar and spreading out radially to the north and south. This
strongly resembles the inner part of a dipole field projected onto the
plane of the sky. The field lines must close, presumably at large
radius.  However, in Figure~\ref{rmpib} we are seeing only the
component perpendicular to the line of sight.  The relatively low
percentage polarization even at very high frequencies of about 25\,\%
- intrinsically it should be about 70\,\% - cannot be explained by
depolarization caused by Faraday rotation, because at high frequencies
this effect is negligible: a RM of 400 rad/m$^2$, the highest we have
detected, rotates the polarization angle by a mere $20^{\circ}$ at
10550~MHz. Hence, the depolarization must be caused by the
superposition of polarized emission generated at different points
along the line of sight with intrinsically different polarization
angles, averaging to a low net polarized fraction.  In a pure dipole
field the component perpendicular to the line of sight, which
determines the observed polarization angle, does not change
significantly along different lines of sight through the nebula.
There must be an additional magnetic field present that also creates
synchrotron emission but has an intrinsic magnetic field configuration
perpendicular to, or at least very different from, the dipole field. A
toroidal field component wrapped around the axis of the dipole would
match that description. Toroidal field structures have been observed
in Vela~X \citep{dods03} and in the PWN in SNR G106.3$+$2.7
\citep{koth06a}, although most other pulsar wind nebulae 
show only an elongated
radial field, indicating the presence of a dipole.  The patches
of polarized emission to the east and west of the pulsar have a
magnetic field structure perpendicular to that expected for a pure
dipole field and could be a hint of a toroidal magnetic field. The two
prominent ``holes'' in polarized intensity would then mark areas where
the polarized emission generated by the dipole component and that
generated by the toroidal component are equally strong but have
orthogonal polarization angles so that the net emission is intense but
unpolarized. Even though both magnetic field configurations may be
present it is clear that the dipole field is the dominant one for
emission, as seen in the magnetic field projected to the plane of the
sky, and for Faraday rotation. If most of the RM was produced by the
toroidal field one half of the nebula would show a negative RM and the
other a positive one.

\begin{figure*}
\centerline{\includegraphics[bb = 45 110 480 510,width=16cm,clip]{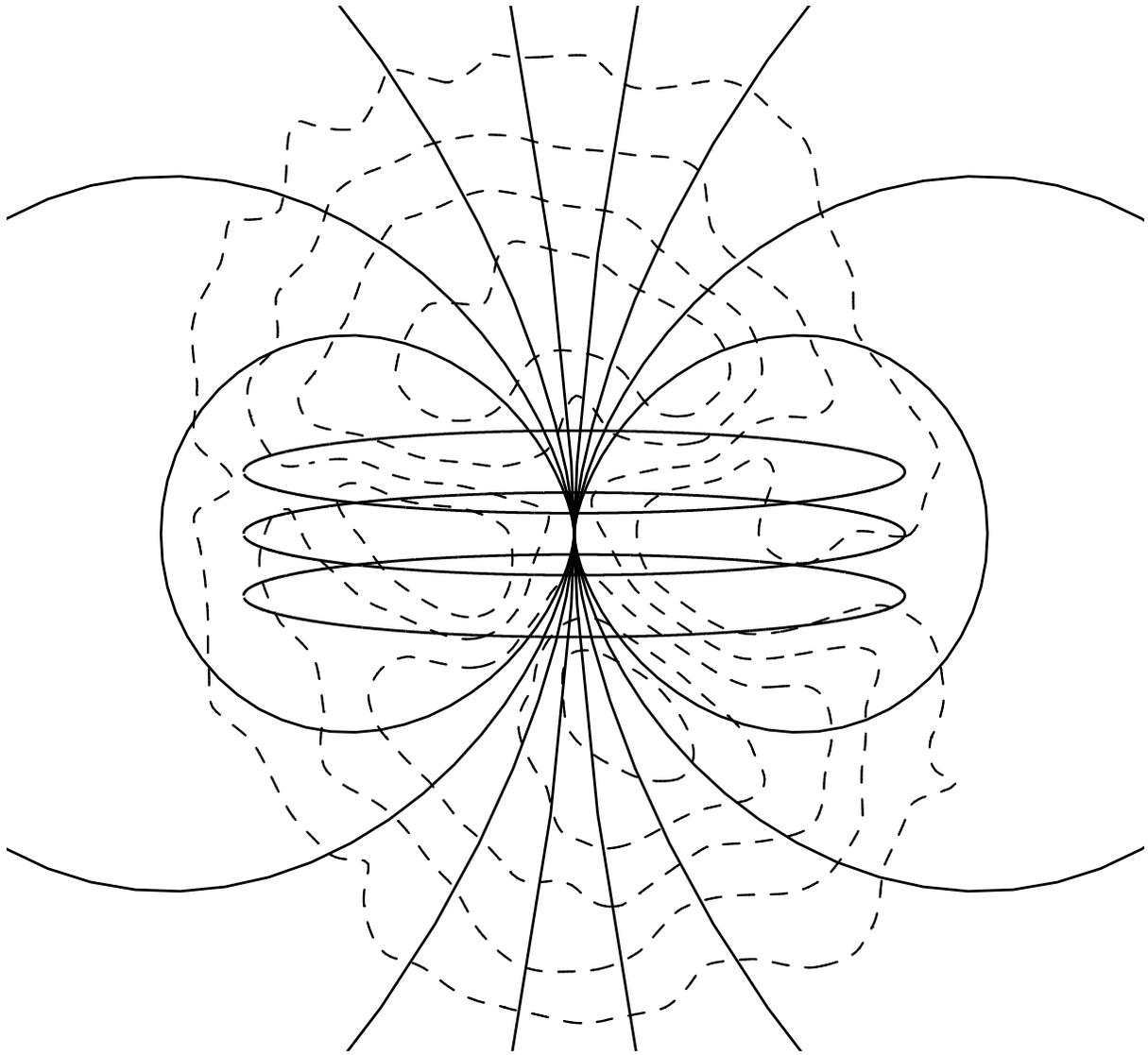}}
   \caption{Sketch of a possible magnetic field configuration inside DA\,495,
   which contains a dipole and a toroidal component. Dashed contours represent
   the polarized emission as observed at 4850~MHz.}
   \label{magsketch}
\end{figure*}

One possible magnetic field configuration that could explain the
radio polarization observations of DA\,495 is shown in
Fig.~\ref{magsketch}.  This is no more than a naive sketch, but it
fits the available observations remarkably well.  A similar
combination of dipole and toroidal field components can explain the
polarization appearance of other PWNe; the polarized emission
structure, the magnetic field vectors projected to the plane of the
sky, the fractional polarization, and the Faraday rotation structure
all depend on the viewing angle and which one of the two components is
dominant \citep{koth06a}.

\subsubsection{The apparent rotation of the polarization structure}

One curious feature of the radio emission from DA\,495 is the
seemingly systematic rotation with frequency of almost the entire
polarized emission structure. At the three higher frequencies we can
see the two prominent lobes rotating from north-east and south-west at
10550~MHz to west and east at 2695~MHz. Near the center the two
polarization holes (or, equivalently, the emission ridge) also rotate
clockwise. At 2695~MHz we can no longer see this rotation near the
center, probably because of the larger beam. This effect must be the
product of depth depolarization in a highly ordered magnetic field
structure.

A possible explanation is the following.  At the
intrinsic (zero wavelength) location of the polarization holes the
magnetic field lines of the toroidal and dipole fields are
perpendicular, and the polarization averages out. Close to the center
of the PWN the dipole field lines are almost all parallel, while the
toroidal field lines change with azimuthal angle within the torus. If
the Faraday rotation of emission generated in the toroidal field
differs from that generated in the dipole field then, at lower
frequencies, this averaging no longer occurs at the ``intrinsic''
location but at a location where the dipole and toroidal field lines
are at some angle other than 90$^{\circ}$. The location of the holes
then moves systematically with frequency, in the sense opposite to
Faraday rotation, as observed.

If the apparent rotation of the polarized structure is a Faraday
rotation effect, then it should show a $\lambda^2$ dependence, and we
can test this. We define $\psi$ as the position angle of a line
through the centers of the lobes ($\psi$ increasing from north to
east).  For 10550~MHz ${\psi} = {45\degr}$ and for 4850~MHz ${\psi}
= {15\degr}$.  Drawing a second line through the polarization holes,
and assuming it is perpendicular to the line through the lobes, we
find ${\psi}={38\degr}$ at 10550~MHz and ${\psi}={10\degr}$ at
4850~MHz.  Averaging these estimates, we derive a ``structural
rotation'' of about $-170$~rad/m$^2$ between 10550 and 4850~MHz and an
intrinsic $\psi$ of $50\degr$. As a test, we extrapolated to
2695~MHz, and predict ${\psi}={-70\degr}$, which is very close to the
$-90\degr$ we observe.  A second test comes from comparison with our
RM map (Figure 5).  Faraday rotation seems to be dominated by the
dipole field, and the RM should then show a maximum or minimum at the
center of the lobes because in those directions the angle between the
line of sight and the field lines is smallest. A line from the RM peak
in the north-eastern lobe to the pulsar has an angle of $55\degr$ with
north, comparable with the intrinsic $\psi$ of $50\degr$ calculated
above.

Since the spin axis of the pulsar is the only axis of symmetry
in the rotating pulsar frame, we can assume that it is parallel to the
axis of the nebula's dipole field, which then would reflect the
rotation-averaged pulsar field. 
We conclude that the spin axis of the pulsar, the main axis
of the toroidal field, and the axis of the dipole magnetic field
component all have an angle $\psi\approx {50\degr}$ projected onto
the plane of the sky. Further, the viewing angle, $\theta$, between the 
line of
sight and the equatorial plane of the dipole (see Fig.~\ref{dipole}), must 
be larger than
$0\degr$. At ${\theta}={0\degr}$ the toroidal field component would
be constant projected onto the plane of the sky and we would not
observe the structural rotation effect.

\subsubsection{Orientation of the dipole field}

In this section we develop a simple model of the magnetic field
configuration in DA\,495. We test the model by computing RM along 
a few selected lines of sight through the nebula, and achieve plausible
agreement with measurements. We have not attempted detailed modeling
of the field.

\begin{figure*}
\centerline{\includegraphics[bb = 55 45 530 515,width=16cm,clip]{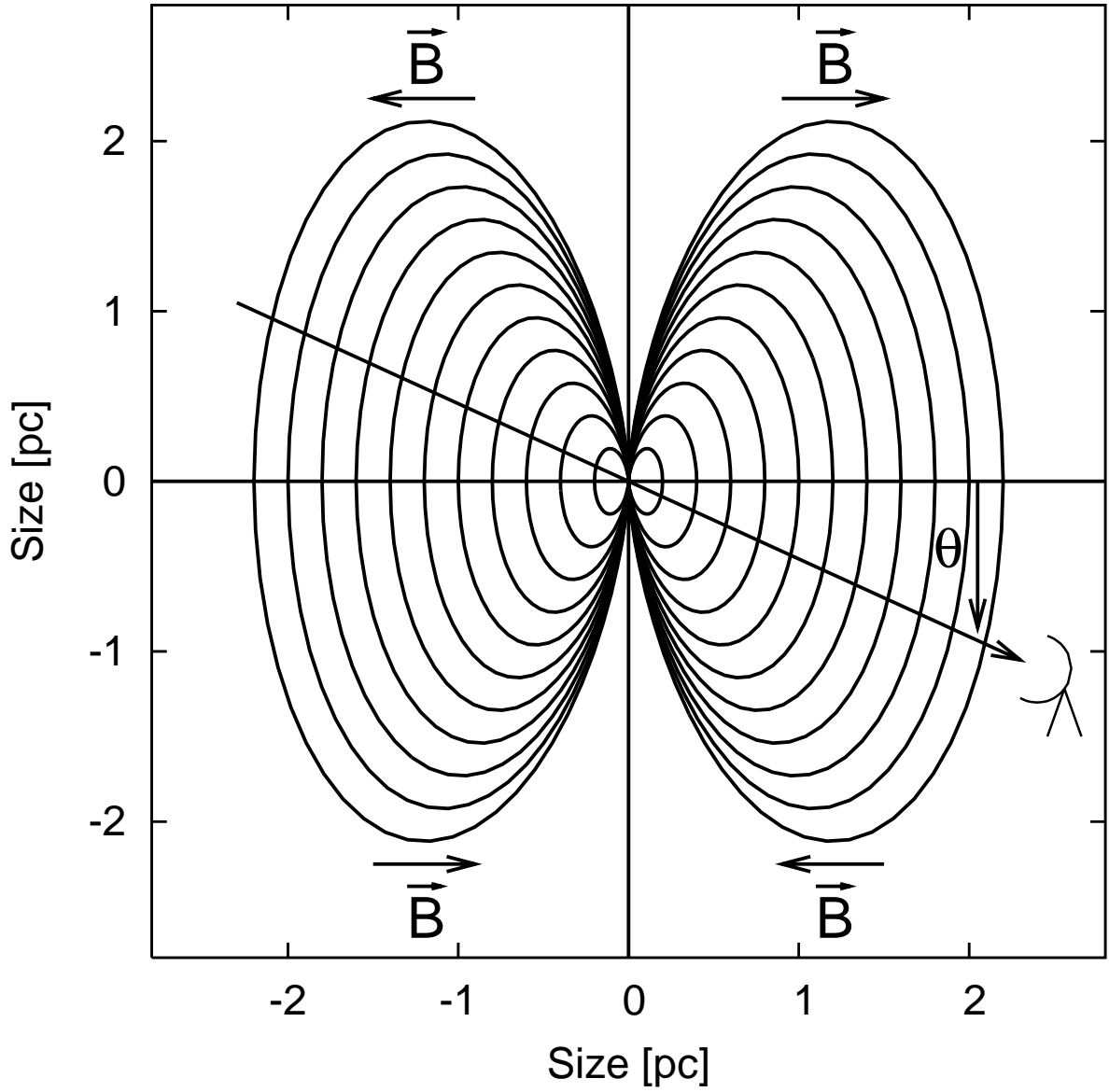}}
   \caption{Sketch of the dipole field configuration adopted as a model
   of one component of the magnetic field in DA\,495. The field lines
   have been assumed to have a dipole configuration that is stretched
   vertically by a factor of 2.5. The direction of the magnetic 
   field $\vec{\rm B}$ is indicated. The spin axis of the pulsar
   should be aligned with the dipole axis. The viewing angle $\theta$
   is defined as shown. In Section 4.4.4 we consider this configuration
   viewed from the right.}
   \label{dipole}
\end{figure*}

At a distance of 1.0~kpc the diameter of DA\,495 is 4.4~pc; we assume
this is the maximum path length through the nebula. We deduced an
internal field of about 1.3~mG from our comparisons with the Vela~X
nebula (Section 4.1 and Table~\ref{DA495_pulsar}) and we assume that
the field is constant at this value throughout the nebula. We know the
projected field distribution (Figure~\ref{rmpib}) and we know that the
dipole field should be the dominating factor for the internal Faraday
rotation. For these calculations we assume a dipole field as displayed
in Figure~\ref{dipole}. 
We confine our attention to the main axis of the dipole, because
any toroidal component will be perpendicular to the line of sight
and will not contribute to RM (and so we ignore the complication of the
toroidal component). Magnetic field strength and synchrotron 
emissivity are assumed
constant throughout the nebula (the evidence for these assumptions 
is presented in Sections 4.2 and 4.3), and the intrinsic polarization angle
is set to be constant at all points along every line of sight through the
nebula. We calculate the electron density $n_e$ within the nebula needed
to generate the observed RM after correction for the foreground.
We make this calculation for various viewing angles, $\theta$. In our 
calculations we are looking from the
right hand side of Fig.~\ref{dipole} into the nebula. We consider we 
have succeeded if we can find an electron density within
the nebula, assumed constant, and a viewing angle which together provide 
RM values reasonably consistent with the observed RM along different
lines of sight. The foreground RM of approximately $-$30 rad/m$^2$
is taken as a measure of the acceptable uncertainty.

Consider first the appearance at ${\theta}={0^{\circ}}$.  In the
northern half the emission generated in the deeper levels first passes
through a magnetic field that is directed away from us on the far side
of the axis and then through a region where it is pointing towards us
on the near side.  Since the emission is generated uniformly along the
line of sight, the bulk of the emission passes through the magnetic
field that is pointing towards us and the resulting rotation measure
is positive. However, in the bottom half the structure is reversed,
and this should lead to a negative RM. We would expect to see a
symmetrical RM map, positive to the north and negative to the south
with peaks of RM of equal absolute value. Now consider the appearance
at an angle ${\theta} > {0^{\circ}}$ as portrayed in
Figure~\ref{dipole}. For lines of sight above the center of the
nebula, positive RM dominates, as above. For lines of sight passing
below the center the emission from the far side of the nebula emerges
with a strong positive RM because the path is long; there is some path
through the near side that contributes negative rotation, and there is
some reduction of net RM, but the path through the near side is
relatively short, so that the final RM value remains strongly
positive, indeed more positive than in the upper lobe. Our observations
do show positive RM throughout, but with a higher RM value on the
north-east lobe. Therefore to represent the observed situation, the
field distribution in Figure~\ref{dipole} must be rotated by
180$^{\circ}$ around the line of sight, so that the dipole is directed
downwards and away from us.

\begin{figure*}
\centerline{\includegraphics[bb = 80 115 825 460,width=16cm,clip]{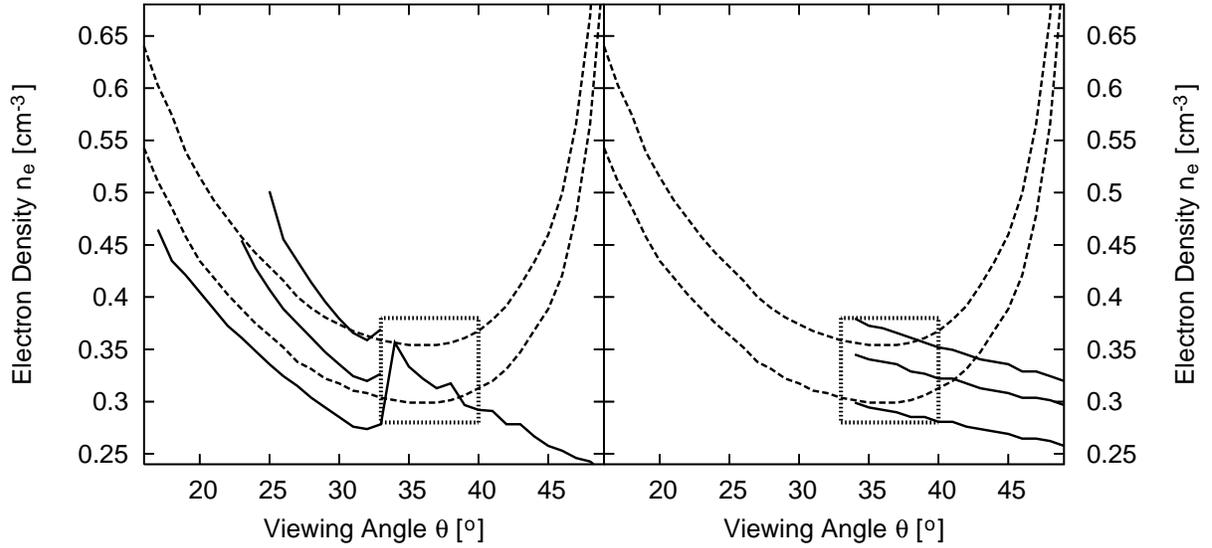}}
   \caption{The required electron density $n_e$ as a function of viewing angle
   $\theta$. The dashed lines in both panels correspond to the line of sight
   through the center of the dipole (Fig.~\ref{dipole}), the lower line
   for the observed rotation measure of +170~rad/m$^2$ and the upper
   line for +200~rad/m$^2$, which is the observed value corrected for
   the foreground RM. The solid lines represent the line of sight through
   the dipole at a distance of 1.2~pc from the center, looking through
   the top of the dipole (left) and the bottom of it (right) as displayed
   in Fig.~\ref{dipole}. The three
   lines were calculated from the top for rotation measures of +350, +300,
   and +250~rad/m$^2$. The small dotted rectangles delineate the range of
   parameter values for which the model can provide a reasonable match
   to observed values of RM.}
   \label{rmcalc}
\end{figure*}

In order to test whether this simple model can predict the observed
properties of DA\,495, we calculated the RM values that would be
observed between 10550 and 4850~MHz for three different paths through
the nebula, through the center and 1.2~pc above and below the
center. At the center we observe a rotation measure of about
170~rad/m$^2$, which amounts to 200~rad/m$^2$ after correction for the
foreground RM. At the RM peak on the north-eastern lobe, 1.2~pc above
the center, we observe about 310~rad/m$^2$ and at the same distance on
the south-western lobe we find about 250~rad/m$^2$. We calculated the
electron density, n$_e$, required to produce these RM values for the
three lines of sight through the nebula for varying viewing angles
$\theta$.  The results are displayed in Fig.~\ref{rmcalc}.  We find
that the model, although approximate, can reproduce the observed RM
values with reasonable values for the parameters.  The small
rectangles identify the region of satisfactory agreement. 
Within the limitations of our assumptions, the electron
density in the nebula lies between 0.3 and 0.4 cm$^{-3}$ and the
viewing angle $\theta$ is between 33$^{\circ}$ and 40$^{\circ}$.

In our model we assumed constant magnetic field throughout the nebula.
Even though the total emissivity and hence the total magnetic field
strength inside the nebula must be relatively constant, we know that
the magnetic field strength in a dipole field decreases with distance
from the centre. We use another simple model, going to the other
extreme, to assess the effect of this assumption: we collapse all the
rotation measure along the line of sight onto the central axis of the
dipole. All the emission generated beyond the axis is rotated and then
added to the emission from the near side. Due to the symmetry of the
dipole, the collapsed RMs at the two positions above and below the
centre used in our calculations would be the same. However, by
changing the viewing angle $\theta$ we would change the pathlength
through the nebula in front of and beyond the central axis.  For
constant emissivity we derive an angle $\theta$ of $10\degr$, in
contrast to the ${\sim}36\degr$ derived above. Under both assumptions,
and also under conditions of varying emissivity within the nebula, the
dipole must be directed away from us towards the south.

\begin{figure*}
\centerline{\includegraphics[bb = 80 115 820 485,width=16cm,clip]{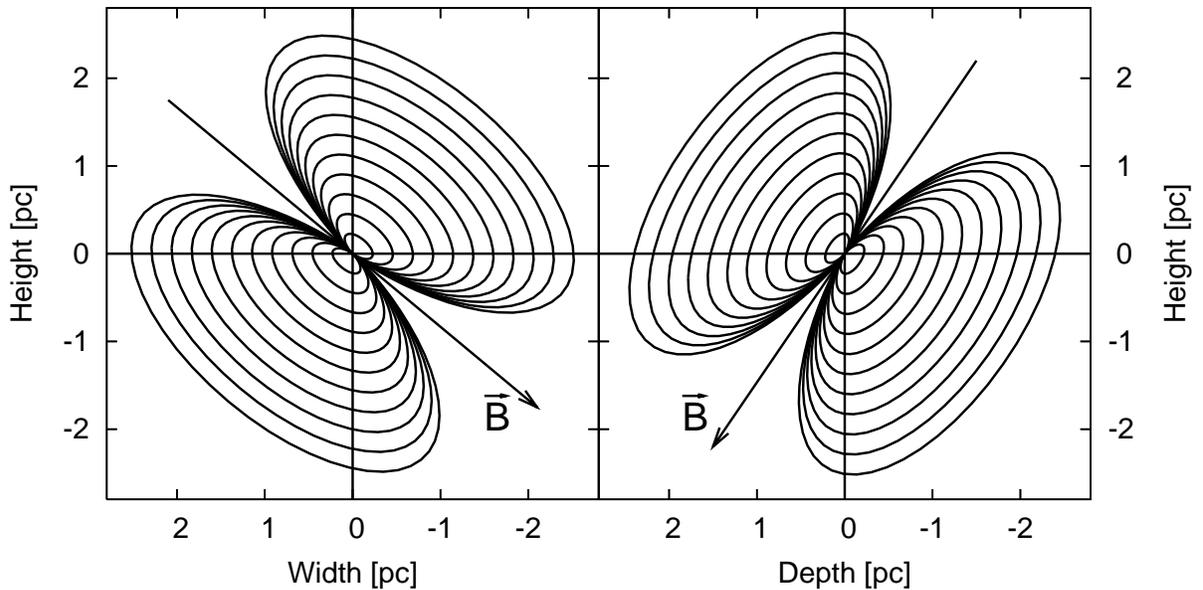}}
   \caption{Orientation of the dipole field inside DA\,495, as determined
   in Sections 4.3 and 4.4. On the left the projection of the dipole field
   onto the plane of the sky is shown. On the right the line of sight
   orientation is indicated with view from the right hand side.}
   \label{oridip}
\end{figure*}

Finally, in Fig.~\ref{oridip} we illustrate the three dimensional
orientation of the dipole field inside DA\,495 as determined in this
section. On the left hand side we display the magnetic field projected
to the plane of the sky. The magnetic field axis has an angle of about
$50\degr$ east of north and the magnetic field is pointing
south-west. On the right hand side the dipole field orientation along
the line of sight is shown, indicating that the dipole is pointing
away from us towards the south-west.

\subsection{Limits on a surrounding shell-type SNR}

Is DA\,495 a pure pulsar wind nebula, like the Crab
Nebula, or does it have a surrounding SNR shell? With our new data we
were able to search for such a shell. None was found, and we can place
sensitive upper limits on any such emission.  The noise in the area
around DA\,495 in our 1420~MHz image, which is the most sensitive we
have for such a search, is about 40~mK rms. A $3\sigma$ signal would
be about 1.0~mJy/beam, and we would need this peak flux density for a
positive detection.  If the inner edge of this shell is at the outer
edge of the PWN and its shell thickness is 10\%, typical for an
adiabatically expanding SNR shell, the upper limit for the total flux
density is about 90~mJy at 1420~MHz. An older remnant would have
higher compression and a thinner shell, resulting in an even lower
flux density limit. The 1~GHz flux density limit is about 100~mJy, using a
spectral index of $\alpha = -0.5$, which is a typical value for a 
shell-type SNR.  This translates to a maximum radio
surface brightness of $\Sigma_{\rm 1~GHz}^{\rm max} = 5.4 \times
10^{-23}$~Watt~m$^{-2}$~Hz$^{-1}$~sr$^{-1}$, slightly
lower than the lowest radio surface-brightness known, that of the SNR
G156.2$+$5.7 which has ${\Sigma_{\rm 1~GHz}}={6 \times
10^{-23}}$~Watt~m$^{-2}$~Hz$^{-1}$~sr$^{-1}$ \citep{reic92}.

The 60$\mu$m image of the DA\,495 field (Fig.~\ref{tp}) shows infrared
emission to the north-east of the SNR. 
We found \ion{H}{1} in the CGPS data, which is likely related to this 
emission, because it exhibits a similar shape. This \ion{H}{1}, however,
is found at velocities near the tangent point, far removed from the systemic 
velocity of DA\,495.
We do not believe that any of this
infrared emission is related to the SNR.

\section{Conclusions}

We have imaged the radio emission from the pulsar wind nebula 
DA\,495 in total intensity
and polarized emission across a range of frequencies from 408 to
10550~MHz.  After correcting for the flux contribution of extraneous
superimposed sources we have demonstrated that there is a break in the
spectrum at about 1.3\,GHz which we can convincingly attribute to
synchrotron cooling. The spectral break at such a low frequency, the
lowest known for any PWN, can only be produced if the pulsar
originally was very energetic, with a high energy loss rate.  We
consider that DA\,495 is an aging pulsar wind nebula which has a pulsar that 
in its 
earlier stages resembled the Crab pulsar or the pulsar in Vela~X. 
The magnetic field inside the nebula is
remarkably well-organized and has the topology of a dipole field
centered on the pulsar; the field probably also has a toroidal
component, wrapped around the dipole axis. The rotation measure is 
high, as would be expected in a
nebula where the magnetic field has the high value of
$\sim$1.3~mG, and the nebula becomes Faraday thick at a very high
frequency.  We have set a very low upper limit for the surface
brightness for any shell of SNR emission surrounding DA\,495. The
absence of a shell implies that the SN explosion occurred in a
low-density environment, and this also explains how the magnetic field
structure has remained so regular to an age of $\sim$20,000 yr.

\section{Acknowledgments}
The Dominion Radio Astrophysical Observatory is a National Facility
operated by the National Research Council. The Canadian Galactic Plane
Survey is a Canadian project with international partners, and is
supported by the Natural Sciences and Engineering Research Council
(NSERC).  This research is based on observations with the 100-m
telescope of the MPIfR at Effelsberg. SSH acknowledges support by 
the Natural Sciences and Engineering Research Council and the 
Canada Research Chairs program. ZA was supported by
NASA grant NRA-99-01-LTSA-070. The VLSS is being carried out by 
the (USA) National Radio Astronomy Observatory (NRAO) and the Naval Research 
Lab. The NRAO is operated by Associated Universities, Inc. and is a 
Facility of the (USA) National Science Foundation.

\appendix

\clearpage

\begin{deluxetable}{lllll}
   \tablewidth{0pc}
   \tablecolumns{5}
   \tablecaption{Observation characteristics\label{obspara}}
   \tablehead{{Frequency [MHz]} & {408} & {1420} 
     & {4850} & {10550}}
   \startdata
Telescope & DRAO & DRAO & Effelsberg & Effelsberg \\
Observation Date & January & January & February & January \\
     &         &         &          & to July \\
     & 2002 & 2002 & 1996  & 1990  \\
Half-power beam ($'$) & $2.9 \times 5.95$ & $0.82 \times 1.75$ & 2.45 & 2.45$^a$ \\
                      & (R.A. $\times$~Dec.) & (R.A. $\times$~Dec.) & & \\
rms noise, $I$, [mJy/beam] & 7 & 0.4 & 0.5 & $<1.5^b$ \\
rms noise, $PI$, [mJy/beam] & & 0.4 & 0.3 & 0.4 \\
Calibrators, $I$, (Flux Density, Jy) & 3C48 (38.9)  & 3C48 (15.7) & 
3C286 (7.5) & 3C286 (4.5) \\
                                    & 3C147 (48.0) & 3C147 (22.0) & & \\
                                    & 3C295 (54.0) & 3C295 (22.1) & & \\
Calibrators $Q, U$ &  & 3C286 & 3C286 & 3C286 \\
Linear Pol. [\%] &  & 9.25 & 11.3 & 11.7 \\
Pol. Angle [$\degr$] &  & 33 & 33 & 33\\
Coverages$^c$ & & & 14 & 5 \\
\enddata
\\$^a$ Original resolution 1\farcm15. Smoothed to 2\farcm45 for
presentation.

$^b$ Estimated. There is no region in the image entirely free of emission.

$^c$ Effelsberg observations only.
\end{deluxetable}

\clearpage

\begin{deluxetable}{cccccc}
   \tablewidth{0pc}
   \tablecolumns{6}
   \tablecaption{Integrated flux densities of DA\,495 \label{fluxes}}
   \tablehead{
\colhead{Frequency [MHz]} & \colhead{Beam} & \colhead{$S_{meas}$ [Jy]} & 
\colhead{$S_{cs-r}$ [Jy]} & \colhead{$S_{HII-r}$ [Jy]} & 
\colhead{reference}}
   \startdata
34.5  & $26' \times~44'$                  & $<35.0$       & $<$24.8        &
& 1 \\
74 & $1\farcm33 \times 1\farcm33$   &               & $<$14          & 
&  5 \\
83    & $0.2^{\circ} \times~5.2^{\circ}$  & $18 \pm~5$    & 13.0 $\pm$~5   &
& 2 \\
111   & $0.15^{\circ} \times~3.9^{\circ}$ & $17 \pm~5$    & 13.0 $\pm$~5   &
& 2 \\
318   & $14' \times~16'$                  & $9.7 \pm~2.2$ & 8.1 $\pm$~2.2  &
& 3 \\
327   & 1\farcm1 $\times$~0\farcm6        &               & $7.8 \pm~1.0$  &
& 4 \\
408   & 2\farcm8 $\times$~5\farcm7        &               & $6.5 \pm~0.6$  &
& 5 \\
430   & 8\farcm4 $\times$~9\farcm6        & $8.7 \pm~4.9$ & $7.5 \pm~4.9$  &
& 3 \\
610.5 & $16' \times~18'$                  & $7.0 \pm~1.5$ & $6.0 \pm~1.5$  &
& 6, 7 \\
1420  & 0\farcm8 $\times$~1\farcm6        &               & $4.0 \pm~0.2$  &
$3.6 \pm~0.3$ & 5 \\
1665  & $22'$                             & $4.4 \pm~0.5$ & $4.0 \pm~0.5$  & 
$3.6 \pm~0.5$ & 7 \\
2695  & 4\farcm3                          &               & $2.7 \pm~0.2$  &
$2.3 \pm~0.3$ & 8 \\
2735  & 4\farcm9 $\times$~5\farcm7        & $2.8 \pm~0.4$ & $2.7 \pm~0.4$  &
$2.3 \pm~0.4$ & 7 \\
4850  & 2\farcm45                         &               & $1.59 \pm~0.1$ &
$1.2 \pm~0.1$ & 5 \\
10550 & 1\farcm15                         &               & $1.12 \pm~0.1$ &
$0.76 \pm~0.18$ & 5 \\
   \enddata
\\Notes: $S_{meas}$ is the measured integrated flux density of
DA\,495. $S_{cs-r}$ is the measured flux density minus the total flux
density of the three brightest compact sources (see text)
appropriately scaled with frequency. For the new data presented in
this paper, and for other high-resolution measurements, this is the
integrated flux density of DA\,495 without the compact
sources. $S_{HII-r}$ is the integrated flux density minus an allowance
for the \ion{H}{2} region which overlaps the SNR (see text).

References: (1) \citet{dwar90}, (2) \citet{kova94}, (3) \citet{dick75}, 
(4) \citet{velu89}, (5) present work, (6) \citet{dick71}, 
(7) \citet{will73}, (8) \citet{reic90}
\end{deluxetable}

\clearpage

\begin{deluxetable}{ccccc}
   \tablewidth{0pc}
   \tablecolumns{5}
   \tablecaption{Flux densities of compact sources [mJy]\label{sourcefd}}
   \tablehead{
\colhead{Frequency, MHz} & \colhead{Source 1} & \colhead{Source 2} & 
\colhead{Source 3} & Reference }  
   \startdata
 74   & 1470 $\pm$~180 & 1200 $\pm$~400 & 900 
$\pm$~300 & 3 \& 10 \\
330   & 613 $\pm$~50 & 711 $\pm$~70 & 328 $\pm$~30 & 1 \\
365   & 557 $\pm$~67 & 739 $\pm$~52 & 294 $\pm$~28 & 2 \\
408   & 650 $\pm$~80 & 504 $\pm$~50 & 200 $\pm$~40 & 3 \\
1420  & 139 $\pm$~5  & 280 $\pm$~8  &  93 $\pm$~5  & 4 \\
1420  & 131 $\pm$~5  & 244 $\pm$~25 &  80 $\pm$~6  & 5 \\
1420  & 134 $\pm$~8  & 288 $\pm$~15 &  83 $\pm$~6  & 3 \\
2695  &              & 160 $\pm$~32 &              & 6 \\
4850  &  50 $\pm$~10 &              &              & 7 \\
4850  &              & 102 $\pm$~15 &              & 8 \\
4850  &              &  89 $\pm$~10 &              & 9 \\
4850  &  58 $\pm$~10 & 108 $\pm$~5  &  36 $\pm$~4  & 3 \\
10550 & 15 $\pm$~4  &  60 $\pm$~10 &  17 $\pm$~5  & 3 \\
   \enddata
\\References: (1) WENSS \citep{reng97}, (2) Texas \citep{doug96}, 
(3) present work, (4) NVSS \citep{cond98}, (5) \citet{land83}, 
(6) \citet{reic90}, (7) MIT-GB \citep{lang90}, (8) 87GB \citep{greg91},
(9) GB6 \citep{greg96}, (10) VLSS \citep{cohe07}
\end{deluxetable}

\begin{deluxetable}{cccc}
   \tablewidth{0pc}
   \tablecolumns{4}
   \tablecaption{Positions and spectra of compact sources\label{sourceps}}
   \tablehead{
 & \colhead{Source 1} & \colhead{Source 2} & \colhead{Source 3}}  
   \startdata
RA(2000)  & 19$^h$ 51$^m$ 52.1$^s$ & 19$^h$ 53$^m$ 3.1$^s$ &
19$^h$ 52$^m$ 40.8$^s$ \\
Dec(2000) & $29^{\circ}~26'~7''$ & $29^{\circ}~10'~36''$ &
$29^{\circ}~35'~47''$ \\
A         & 5.05 $\pm$~0.09 & 4.69 $\pm$~0.07 & 4.58 
$\pm$~0.11 \\
B         & $-$0.92 $\pm$~0.03 & $-$0.72 $\pm$~0.02 &  $-$0.83 
$\pm$~0.04 \\
   \enddata
\\log(S[mJy]) = A + B~log($\nu$[MHz])
\end{deluxetable}

\begin{deluxetable}{ccc}
   \tablewidth{0pc}
   \tablecolumns{3}
   \tablecaption{Integrated polarization properties\label{intpol}}
   \tablehead{
Frequency [MHz] & \colhead{Polarized Intensity [Jy]} &
\colhead{Fractional Polarization [\%]}}
   \startdata
1420  & $0.44 \pm~ 0.02$ & $12 \pm~2$ \\
2695  & $0.59 \pm~ 0.05$ & $26 \pm~4$ \\
4850  & $0.29 \pm~ 0.03$ & $24 \pm~4$ \\
10550 & $0.18 \pm~ 0.02$ & $24 \pm~6$ \\
   \enddata
\\Fractional polarization is computed using integrated flux densities
corrected for compact sources and the superimposed \ion{H}{2}
region. See Table~\ref{fluxes} 
\end{deluxetable}

\begin{deluxetable}{lccc}
   \tablewidth{0pc}
   \tablecolumns{4}
   \tablecaption{Breaks in the spectra of pulsar wind nebulae
\label{break_freq}}
   \tablehead{
\colhead{SNR} & \colhead{Break Frequency} & \colhead{injected/cooling$^a$} &
\colhead{reference}}
   \startdata
Crab Nebula & 40~keV~\& 1000~\AA & {\it{i}} & \citet{wolt97} \\
Crab Nebula & 14000 GHz & {\it{c}} & \citet{stro92} \\
W44 & 8000 GHz & {\it{c}} & \citet{petr02} \\
Vela X & 100 GHz & {\it{c}} & \citet{weil80} \\
G29.7$-$0.3 & 55 GHz & {\it{i}} & \citet{bock05} \\
3C~58 & 50 GHz & {\it{i}} & \citet{salt89,gree94} \\
G21.5$-$0.9 & $\sim 30$~GHz & {\it ?} & \citet{salt89} \\
G16.7+0.1 & 26 GHz & {\it{i}} & \citet{bock05} \\
CTB 87 & 10 GHz & {\it{c}} & \citet{mors87} \\
G27.8$+$0.6 & 5-10~GHz & {\it{c}} & \citet{reic84} \\
G106.3+2.7 & 4.5 GHz & {\it{c}} & \citet{koth06a} \\
DA 495 & 1.3 GHz & {\it{c}} & this paper \\ 
   \enddata
\\ $^a$ {\it{i}} denotes spectral break due to a break in the injected electron
spectrum; {\it{c}} denotes spectral break due to synchrotron cooling
\end{deluxetable}

\begin{deluxetable}{lcccccc}
   \tablewidth{0pc}
   \tablecolumns{7}
   \tablecaption{Present day and birth properties of the six pulsars for which 
   a reliable age can be found.
For a description of individual parameters see Section 4.2.
\label{pulsars}}
   \tablehead{
\colhead{SNR} & \colhead{Pulsar} & \colhead{$\tau$ [yr]} &
\colhead{$\dot{E}$ [erg/s]} & \colhead{$\tau_0$ [yr]} &
\colhead{$\dot{E}_0$ [erg/s]} & \colhead{$t$ [yr]}}
   \startdata
Crab Nebula & B0531+21     & 1270  & $4.4\times 10^{38}$ &
320         & $1.0\times 10^{40}$ & 950 \\
3C~58       & J0205+6449   & 5370  & $2.7\times 10^{37}$ &
4550        & $3.8\times 10^{37}$ & 820 \\
G11.2$-$0.3 & J1811$-$1925 & 23300 & $6.4\times 10^{36}$ &
21680       & $7.4\times 10^{36}$ & 1620 \\
Vela X ($\beta = 2.0$) & B0833-45     & 11300 & $6.9\times 10^{36}$ &
110         & $7.3\times 10^{40}$ & 11200 \\
Vela X ($\beta = 6.0$) & B0833-45     & 11300 & $6.9\times 10^{36}$ &
3000         & $2.0\times 10^{40}$ & 8300 \\
G106.3+2.7  & J2229+6114   & 10460 & $2.2\times 10^{37}$ &
6560        & $5.6\times 10^{37}$ & 3900 \\
W~44        & B1853+01     & 20300 & $4.3\times 10^{35}$ &
13300       & $1.0\times 10^{36}$ & 7000 \\
   \enddata
\end{deluxetable}

\begin{deluxetable}{lccccc}
   \tablewidth{0pc}
   \tablecolumns{6}
   \tablecaption{Properties of DA 495 extrapolated from birth
   properties of the six pulsars in Table~\ref{pulsars}.
\label{DA495_pulsar}}
   \tablehead{
\colhead{Basis} & \colhead{$t_{\rm DA\,495}$ [yr]} & \colhead{$\tau_{\rm DA\,495}$ 
[yr]} & \colhead{$B_{\rm req}$ [mG]} & \colhead{$E_{\rm tot}$ [erg]} &
\colhead{$B_{\rm max}$ [mG]}}
   \startdata
Crab Nebula ($\beta = 2.0$) & 100900 & 101200 & 0.45 & $1.0\times 10^{50}$ & 0.98 \\
Crab Nebula ($\beta = 2.3$) & 47600 & 47900 & 0.74 & $7.8\times 10^{49}$ & 0.86 \\
3C~58       & 84100  & 88700  & 0.51 & $5.2\times 10^{48}$ & 0.22 \\
G11.2$-$0.3 & 165000 & 186000 & 0.32 & $4.5\times 10^{48}$ & 0.22 \\
Vela X ($\beta = 2.0$)  & 93900  & 94000  & 0.47 & $2.5\times 10^{50}$ & 1.56 \\
Vela X ($\beta = 6.0$)  & 20000  & 23000  & 1.32 & $3.8\times 10^{50}$ & 1.91 \\
G106.3+2.7  & 149000  & 155000 & 0.35 & $1.1\times 10^{49}$ & 0.32 \\
W~44        & 28800  & 42100  & 1.03 & $2.3\times 10^{47}$ & 0.05 \\
   \enddata
\end{deluxetable}
\clearpage

\begin{thebibliography}{}


\bibitem[Arzoumanian et al.(2004)]{arzo04} Arzoumanian Z., Safi-Harb
S., Landecker T.L., Kothes R., 2004, \apjl, 610, L101
	
\bibitem[Arzoumanian et al.(2008)]{arzo08} Arzoumanian Z., Safi-Harb
S., Landecker T.L., Kothes R., Camilo, F., 2008, \apj, accepted,
ArXiv Astrophysics e-prints arXiv:0806.3766
	

\bibitem[Blondin et al.(2001)]{blon01} Blondin J.M., Chevalier R.A.,
   Frierson D.M., 2001, \apj, 563, 806

\bibitem[Bock \& Gaensler(2005)]{bock05} Bock D. C.-J., Gaensler, B.M.,
2005, \apj, 626, 343

\bibitem[Burn(1966)]{burn66} Burn, B.J., 1966, \mnras, 133, 67

\bibitem[Cao et al.(1997)]{cao97} Cao, Y., Terebey, S., Prince, T.A., 
   Beichman, C.A., 1997, \apjs, 111, 387

\bibitem[Chevalier(2000)]{chev00} Chevalier R.A., 2000, \apj, 539, L45

\bibitem[Cohen et al.(2007)]{cohe07} Cohen A.S., Lane W.M., Cotton W.D., 
   Kassim N.E., Lazio T.J.W., Perley R.A., Condon J.J., Erickson W.C.,
   2007, \aj, 134, 1245

\bibitem[Condon et al.(1998)]{cond98} Condon J.J., Cotton W.D.,
Greisen E.W., Yin Q.F., Perley R.A., Taylor G.B., Broderick J.J., 1998,
\aj, 115 1693

\bibitem[Condon(1984)]{cond84} Condon J.J., 1984, \apj, 287, 461

\bibitem[Cordes \& Lazio(2002)]{cord02}
Cordes, J.~M. \& Lazio, T.~J.~W., 2002, ArXiv Astrophysics e-prints,
astro-ph/0207156

\bibitem[Dickel \& DeNoyer(1975)]{dick75} Dickel J.R., DeNoyer L.K.,
1975, \aj, 80, 437

\bibitem[Dickel et al.(1971)]{dick71} Dickel J.R., Webber J.C., Yang
K.S, Staff, 1971, \aj, 76, 294

\bibitem[Dodson et al.(2003)]{dods03} Dodson R., Lewis D., McConnell D.,
    Deshpande A. A., 2003, \mnras, 343, 116

\bibitem[Douglas et al.(1996)]{doug96} Douglas J.N., Bash F.N.,
Bozyan F.A., Torrence G.W., Wolfe C., 1996, \aj, 111, 1945

\bibitem[Duncan et al.(1999)]{dunc99} Duncan, A.R., Reich P., Reich W.
F\"urst E., 1999, \aap, 350, 447

\bibitem[Dwarakanath \& Udaya Shankar(1990)]{dwar90} Dwarakanath
K.S., Udaya Shankar N., 1990, J. Astrophys. Astr., 11, 323

\bibitem[Emerson \& Gr\"ave (1988)]{emer88} Emerson D.T., Gr\"ave R., 
   1988, \aap 190, 353

\bibitem[Emerson et al.(1979)]{emer79} Emerson D.T., Klein U., Haslam C.G.T.
   1979, \aap, 76, 92
	
\bibitem[Fesen et al.(1992)]{fese92} Fesen, R.A., Martin, C.L., Shull, J.M.,
   1992, \apj, 399, 599

\bibitem[Foster \& MacWilliams(2006)]{fost05} Foster, T., MacWilliams, J.
   2006, \apj, in press (May, 2006, \apj, 642)

\bibitem[F\"urst \& Reich (1986)]{furs86} F\"urst, E., Reich, W. 
   1986, \aap 163, 185
   

\bibitem[Green(1987)]{gree87} Green D.A., 1987, \mnras, 225, L11

\bibitem[Green(1994)]{gree94} Green, D.A., 1994, \apjs, 90, 871

\bibitem[Green(2004)]{gree04} Green D.A., 2004, Bulletin
   of the Astronomical Society of India, 32, 335
   (also available on the World Wide Web at 
   http://www.mrao.cam.ac.uk/surveys/snrs)

\bibitem[Gregory \& Condon(1991)]{greg91} Gregory P.C., Condon J.J.,
1991, \apjs, 75, 1011

\bibitem[Gregory et al.(1996)]{greg96} Gregory P.C., Scott W.K.,
Douglas K., Condon J.J., 1996, \apjs, 103, 427

\bibitem[Haslam et al.(1982)]{hasl82} Haslam C.G.T., Stoffel H., Salter
   C.J., Wilson W.E., 1982, \aaps 47, 1

\bibitem[Hill(1967)]{hill67} Hill E.R., 1967, AuJPh, 29, 29

\bibitem[Kothes et al.(2004)]{koth04} Kothes R., Landecker T.L.,
   Wolleben M., 2004, \apj, 607, 855

\bibitem[Kothes et al.(2006a)]{koth06a} Kothes, R., Reich, W., 
   Uyan{\i}ker, B., 2006a, \apj, 638, 225
   
\bibitem[Kothes et al.(2006b)]{koth06b} Kothes, R., Fedotov, K., Foster,
   T.J., Uyan{\i}ker, B., 2006b, \aap, 457, 1081

\bibitem[Kovalenko et al.(1994)]{kova94} Kovalenko A.V., Pynzar A.V.,
   Udal'tsov V.A., 1994, Astr. Rep., 38, 95

\bibitem[Landecker \& Caswell(1983)]{land83} Landecker, T.L., Caswell,
   J.L., 1983, \aj, 88, 1810

\bibitem[Landecker et al.(1990)]{land90} Landecker T.L., Clutton-Brock M,
Purton C.R., 1990, \aap, 232, 207



\bibitem[Langston et al.(1990)]{lang90} Langston G.I., Heflin M.B.,
Conner S.R., Lehar J., Carilli C.L., Burke B.F., 1990, \apjs, 72, 621

\bibitem[Lyne et al.(1996)]{lyne96} Lyne A.G., Pritchard R.S.,
   Graham-Smith F., Camillo F., 1996, Nature 381, 497

\bibitem[Lyne et al.(1988)]{lyne88} Lyne A.G., Pritchard R.S., Smith F.G.,
   1988, \mnras, 233, 667
   
\bibitem[Lyne \& Smith(1989)]{lyne89} Lyne A.G., Smith F.G., 1989, 
\mnras, 237, 533

\bibitem[Manchester et al.(2005)]{manc05} Manchester, R. N., Hobbs, G. B., 
   Teoh, A., Hobbs, M., \aj, 129, 1993

\bibitem[Morsi \& Reich(1986)]{mors86} Morsi H.W., Reich W., 
   1986, \aap, 163, 313

\bibitem[Morsi \& Reich(1987)]{mors87} Morsi H.W., Reich W., 
   1987, \aap, 69, 533

\bibitem[Pacini \& Salvati(1973)]{paci73} Pacini F., Salvati M.,
   1973, \apj, 186, 249

\bibitem[Petre et al.(2002)]{petr02} Petre, R., Kuntz, K.D., Shelton, R.L.,
2002, \apj, 579, 404


\bibitem[Reich et al.(1984)]{reic84} Reich, W., F\"urst, E., Sofue, Y.,
    1984, \aap, 133, L4

\bibitem[Reich et al.(1990)]{reic90} Reich W., F\"urst E., Reich P., Reif
    K., 1990, \aaps, 85, 633
	
\bibitem[Reich et al.(1992)]{reic92} Reich W., F\"urst E., 
   Arnal E.M., 1992, \aap, 256, 214

\bibitem[Rengelink et al.(1997)]{reng97} Rengelink R.B., Tang Y., de
Bruyn A.G., Miley G.K., Bremer M.N., Roettgering H.J.A., Bremer
M.A.R., 1997, \aaps, 124, 259

\bibitem[Salter et al.(1989)]{salt89} Salter C.J., Reynolds S.P., Hogg D.E.,
    Payne J.M., Rhodes P.J., 1989, \apj, 338, 171
    
\bibitem[Simard-Normandin \& Kronberg(1979)]{sima79} Simard-Normandin M., 
Kronberg P.P., 1979, Nature, 279, 115
	
\bibitem[Sokoloff et al.(1998)]{soko98} Sokoloff, D.D., Bykov, A.A.,
Shukurov, A., Berkhuijsen, E.M., Beck, R., \& Poezd, A.D., 1998,
MNRAS, 299, 189

\bibitem[Stephenson \& Green(2002)]{step02} Stephenson, F.R., 
   Green, D.A., 2002, Historical Supernovae and their Remnants, 
   Oxford University Press

\bibitem[Strom \& Greidanus(1992)]{stro92} Strom, R.G., Greidanus, H.,
1992, Nature, 358, 654

\bibitem[Taylor \& Cordes(1993)]{tayl93} Taylor J.H., Cordes J.M.,
   1993, \apj, 411, 674

\bibitem[Taylor et al.(2003)]{tayl03} Taylor A.R., Gibson S.J., Peracaula,
   M. et al.,
   2003, \aj, 124, 3145

\bibitem[Uyan{\i}ker et al.(2003)]{uyan03} Uyan{\i}ker B., Landecker T.L.,
   Gray A.D., Kothes R., 2003, \apj, 585, 785
	
\bibitem[Velusamy et al.(1989)]{velu89} Velusamy T., Becker R.H.,
Goss W.M., Helfand D.J., 1989, J. Astrophys. Astr., 10, 161

\bibitem[Weiler \& Panagia(1980)]{weil80} Weiler K.W., Panagia N.,
   1980, \aap, 90, 269

\bibitem[Willis(1973)]{will73} Willis A.G., 1973, \aap, 26, 237

\bibitem[Woltjer et al.(1997)]{wolt97} Woltjer L., Salvati M., 
   Pacini F., Bandiera R., 1997, \aap, 325, 295

\bibitem[Xiao et al.(2008)]{xiao08} Xiao L., F\"urst E., Reich W., Han J.L.,
2008, \aap, 482, 783

\end{thebibliography}
\end{document}